\documentclass[fleqn,usenatbib]{mnras}

\usepackage{newtxtext,newtxmath}
\usepackage{threeparttable}

\usepackage[T1]{fontenc}

\DeclareRobustCommand{\VAN}[3]{#2}
\let\VANthebibliography\thebibliography
\def\thebibliography{\DeclareRobustCommand{\VAN}[3]{##3}\VANthebibliography}

\usepackage{graphicx}	
\usepackage{amsmath}	
\usepackage{orcidlink}

\title[Multiple collisions in N59 bubble]{Multiple collisions in N59 bubble: Sequential cloud-cloud collisions}

\author[E. Chen et al.]{
En Chen,$^{1}$\thanks{E-mail: chenen@gzhu.edu.cn}\orcidlink{0009-0002-8449-1734}
Xi Chen,$^{1}$\thanks{E-mail: chenxi@gzhu.edu.cn}\orcidlink{0000-0002-5435-925X}
Xuepeng Chen,$^{2}$\orcidlink{0000-0003-3151-8964}
Min Fang$^{2}$\orcidlink{0000-0001-8060-1321}
and Qianru He$^{2,3}$\orcidlink{0000-0002-5077-9599}
\\
$^{1}$Center for Astrophysics, Guangzhou University, Guangzhou, 510006, People's Republic of China\\
$^{2}$Purple Mountain Observatory $\&$ Key Laboratory of Radio Astronomy, No.10 Yuanhua Road, Nanjing, 210034, People's Republic of China\\
$^{3}$University of Science and technology of China, No.96 Jinzhai Road, Hefei, 230026, People's Republic of China
}

\date{Accepted 2024 October 16. Received 2024 September 27; in original form 2024 July 11. }

\pubyear{2015}

\begin{document}
\label{firstpage}
\pagerange{\pageref{firstpage}--\pageref{lastpage}}
\maketitle

\begin{abstract}

We report that the gas components in the N59 bubble suffered from sequential multiple cloud-cloud collision (CCC) processes. 
The molecular gas in the N59 bubble can be decomposed into four velocity components, namely Cloud A [95, 108] km s$^{-1}$, Cloud B [86, 95] km s$^{-1}$, Cloud C [79, 86] km s$^{-1}$ and Cloud D [65, 79] km s$^{-1}$. 
Four CCC processes occurred among these four velocity components, i.e., Cloud A vs. Cloud B, Cloud A vs. Cloud C, Cloud C vs. Cloud D, and Cloud A vs. Cloud D. 
Using $Spitzer$ MIR and UKIDSS NIR photometric point source catalogs, we identified 514 YSO candidates clustered in 13 YSO groups, and most of them ($\sim60\%$) were located at the colliding interfaces, indicating that they were mainly triggered by these four CCC processes. 
We also found that these four collisions occurred in a time sequential order: the earliest and most violent collision occurred between Cloud A and Cloud D about 2 Myr ago, then Cloud B collided with Cloud A about 1 Myr ago, and finally, Cloud C collided with Clouds A and D simultaneously about 0.4 Myr ago.

\end{abstract}




\section{Introduction} \label{sec:intro}

Star formation, especially massive star (> 8 M$_{\odot}$) formation, is one of the most important astrophysical processes controlling the evolution of galaxies. 
However, the massive star formation process is a long-standing mystery due to their rarity, short lifetime, and deep embedded in their parent molecular clouds. 
Recent studies \citep{Gong2017, Torii2011,Torii2015,Torii2017, Hayashi2018} propose that cloud-cloud collision (CCC) processes can trigger star formation in a variety of mass ranges, from low-mass young stellar objects (YSOs) to massive OB-type stars. 
\cite{Fukui2021} found that almost all massive star forming regions in the Milky Way, such as M42, NGC 6334, etc., are triggered by CCC processes. 
Collisions of these giant molecular clouds (GMCs) tend to form massive star clusters or small starbursts, suggesting that CCC is an important mechanism for massive star formation. 
The review of \cite{Fukui2021} summarizes three observational signatures of CCCs: a) bridge feature, b) complementary spatial distribution, and c) U-shape cavity. 
These signatures are useful guidance to search more CCC candidates in spectroscopic observations. 

The difficulty of massive star formation lies in achieving its initial conditions (ultra-high density) and balancing its strong feedback (radiation pressure vs. gravitational potential). 
The CCC process, as a dynamic feedback, works well on generating the ultra-high density conditions \citep{Takahira2018}. 
A picture of CCC in \cite{Habe1992} shows that the shock waves generated by the supersonic collision can effectively compress gas to form a dense layer, which is a hotbed for massive star formation. 
Recent hydrodynamical simulations \citep{Takahira2014, Takahira2018} reinforce this picture of CCC that collisions with different initial conditions (i.e., size, mass, collision speeds) can produce dense cores with power-law mass functions. 
\cite{Fukui2021} found that the number of O-type stars triggered by CCC is related to its column density, and only when the column density is greater than $10^{22}$ cm$^{-2}$ can an O-type star be formed. 
Therefore, CCC is a very efficient factory for massive star formation. 

The mid-infrared (MIR) bubble N59 was initially documented by \cite{Churchwell2006}, which records bubble structures with mid-infrared emission in the first ($l=10^{\circ}-65^{\circ},\ |b|\leq1$) and the fourth ($l=295^{\circ}-350^{\circ},\ |b|\leq1$) Galactic quadrant of the inner Galactic plane. 
Subsequently, N59 was also documented in the $Spitzer$ Galactic bubbles sample from \cite{Hattori2016} and catalog of AKARI IR bubbles from \cite{Hanaoka2019}. 
The N59 bubble is located in the first Galactic quadrant with coordinates of $(l,b)=(33^{\circ}.071,-0^{\circ}.075)$ and a distance of about $4.66\sim5.6$ kpc. 
The location is close to the famous meeting point of Galactic bar and spiral arm (the Scutum-Centaurus arm), where a mini-starburst W43 (coordinates of $(l, b)=(30^{\circ}.8, -0^{\circ}.05)$ and a distance of 5.6 kpc) exists. 
\cite{Kohno2021} reported that the formation of this mini-starburst may be caused by the collision of gas flows from the bar and the arm. Thus, the similar N59 may have the same origin as W43. 

\cite{Deharveng2010} reported that the kinematic distance of N59 is about 5.6 kpc. Since N59 is nearly located toward the Galactic Center direction, the measurement uncertainty of kinematic distance is very large. 
Using the more accurate Gaia parallax measurements (\citealt{Gaia2016, Gaia2021}), \cite{Paulson2024} obtained an mean distance of 4.66 ± 0.70 kpc for the N59 bubble. 
Therefore, in this paper, we take the result of \cite{Paulson2024} of 4.66 kpc as the distance of the N59 bubble.

\begin{figure*}
	\includegraphics[width=\linewidth]{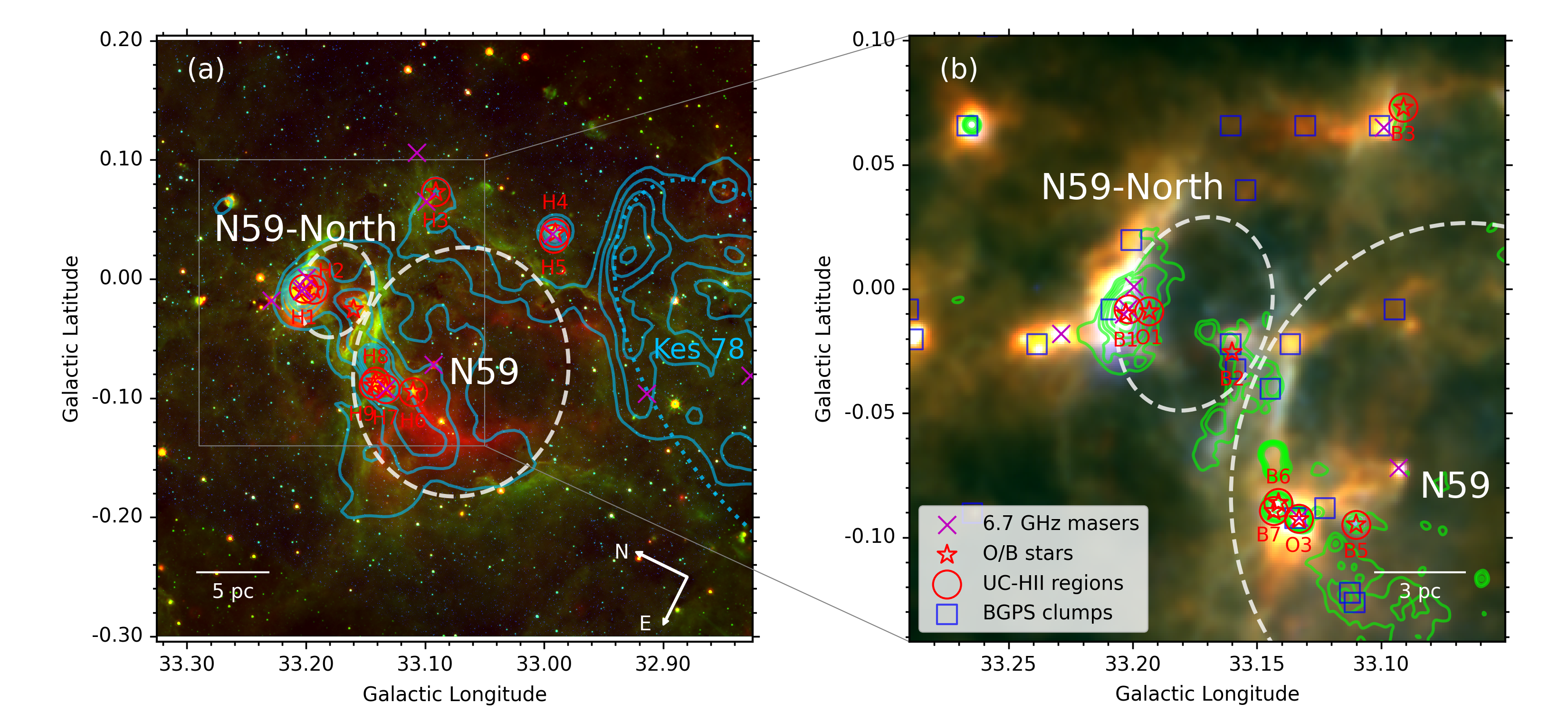}
    \caption{RGB-composite image of the N59 bubble. (a) R, G and B colors indicate $Spitzer$-MIPS 24 $\mu$m, $Spitzer$-IRAC 8 $\mu$m and UKIDSS-K band 2.2 $\mu$m, respectively. The N59 and N59-North bubbles are depicted by white dashed circles. The blue contours indicate VGPS radio continuum emission, which start from 20 to 40 K with steps of 5 K. The blue dotted ellipse fits the SNR Kes 78. (b) Zoom-in region of the N59-North bubble, with its R, G and B colors encoded in $Herschel$ 250, 160, and 70 $\mu$m, respectively. The green contours indicate MAGPIS 20 cm radio continuum emission, with contour levels of 0.001, 0.0012, 0.0015, 0.002, 0.003 and 0.005 Jy/beam. The blue squares indicate the Bolocam clumps at 1.1 mm. To all panels, the red circles indicate the \textsc{Hii} regions. The magenta crosses indicate the 6.7 GHz methanol masers. And the red open star symbols indicate the O/B-type stars. }
    \label{fig:rgb}
\end{figure*}

\setlength{\tabcolsep}{5pt}
\begin{table*}
    \centering
    \caption{Properties of \textsc{Hii} regions in N59. }
    \begin{threeparttable}
    \begin{tabular}{ccccclcccccc}
    \hline
       \textsc{   Hii} & Source  &  $l$  &  $b$  &  Object  & Exciting  & Maser & $\nu$ & $S_{\nu}$ & $lg(\dot{N}_{LyC})$ & $R_s$ &  Age\\
       region  &  name\tnote{a} &  ($^{\circ}$) & ($^{\circ}$) &  type\tnote{b} & O/B stars\tnote{c} & detection & (GHz) & (Jy) & (s$^{-1}$) & (pc) & (Myr)\\
    \hline
       H1 & IRAS 18494+0010  & 33.2017 & -0.0082 & UC  & B1 (B0V) & 2 & 1.49 & 0.075 & 47.11 & 0.04 & 0.002\\
       H2 & KB94  & 33.1934 & -0.0090 &  E  & O1 (O9.5V)  & ... & 1.49 & 0.342 & 47.77 & 0.10 & 2.032\\
       H3 & HRDS G033.091+00.073  & 33.0910 & 0.0730 & UC & B3 (B0.5V)  & 1 & 1.49 & 0.055 & 46.97 & 0.03 & 0.002\\
       H4 & CORNISH G032.9906+00.0385  & 32.9906 & 0.0385  & UC & B4 (B1V)  & ... & 1.49 & 0.009 & 46.19 & 0.02 & 0.001\\
       H5 & WHX2007 18488+0000SE  & 32.9917 & 0.0338 &  UC &  O2 (O9.5V)  & 1 & 1.49 & 0.194 & 47.52 & 0.05 & 0.003\\
       H6 & WBH2005 G033.110-0.095  & 33.1100 & -0.0950 &  UC  & B5 (B1V) & ... & 1.49 & 0.008 & 46.13 & 0.02 & 0.001\\
       H7 & CORNISH G033.1328-00.0923  & 33.1331 & -0.0925 & UC & O3 (O9.5V) & 1 & 8.49 & 0.471 & 47.98 & 0.07 & 0.004\\
       H8 & GAL 033.12-00.08  & 33.1414 & -0.0862  & UC & B6 (B0.5V) & ... & 8.49 & 0.058 & 47.07 & 0.04 & 0.002\\
       H9 & GAL 033.13-00.09  & 33.1432  & -0.0894  & UC & B7 (B0.5V) & ... & 8.49 & 0.047 & 46.98 & 0.03 & 0.002\\
    \hline
    \end{tabular}
    \begin{tablenotes}
        \tiny
        \item[a]  Source names in literature. IRAS 18494+0010 (H1) and GAL 033.13-00.09 (H9) are from the UC \textsc{Hii} region catalog of \cite{Bronfman1996}. KB94 (H2) is a radio \textsc{Hii} region documented by \cite{Lockman1989}. HRDS G033.091+00.073 (H3) is from the source catalog of the Green Bank Telescope (GBT) \textsc{Hii} region discovery survey (HRDS) of \cite{Anderson2011}. CORNISH G032.9906+00.0385 (H4) is from the hypercompact (HC) \textsc{Hii} catalog operated by \cite{Yang2019}. WHX2007 18488+0000SE (H5) is a UC \textsc{Hii} region source from \cite{Wu2007}. WBH2005 G033.110-0.095 (H6) is a UC \textsc{Hii} region source from \cite{Urquhart2009}. CORNISH G033.1328-00.0923 (H7) is from the UC \textsc{Hii} region catalog of \cite{Kalcheva2018}. GAL 033.12-00.08 (H8) is a radio \textsc{Hii} region source from \cite{Lockman1989}. 
        \item[b]  Object type of the \textsc{Hii} regions. "UC" indicate the ultra-compact \textsc{Hii} region, while "E" indicate the evolved expanding \textsc{Hii} region. 
        \item[c]  The exciting O/B-type stars (with their spectral type, \citealt{Paulson2024} and \citealt{Smith2002}) in the \textsc{Hii} regions. The ZAMS indicate the zero age main sequence. 
    \end{tablenotes}
    \end{threeparttable}
    \label{tab:HII}
\end{table*}

Figure \ref{fig:rgb}(a) shows the RGB-composite image of the N59 bubble, in which R, G and B color indicate $Spitzer$-MIPS 24 $\mu$m, $Spitzer$-IRAC 8 $\mu$m and UKIDSS-K band 2.2 $\mu$m, respectively. 
Two ring-like bubbles, namely the N59 bubble and N59-North bubble, are very prominent within this map. The N59 bubble has a diameter of about 16 pc, and the N59-North bubble has a diameter of 5.7 pc. 
The ring structure of the N59 bubble is broken toward its south, where a strong radio continuum emission is visible. This anomalous radio continuum emission comes mainly from the famous supernova remnant (SNR) Kes 78 \citep{Kesteven1968}. 
\cite{Zhou2011} reported that Kes 78 is a very young SNR with an age of about 6 kyr, associated with the molecular clouds at a systemic velocity of 81 km s$^{-1}$. 
Since the molecular cloud velocity associated with N59 is between 65 and 110 km s$^{-1}$ \citep{Paulson2024}, this SNR is also associated with N59. 
We retrieved 9 \textsc{Hii} regions from the online tool $SIMBAD$\footnote{\url{http://simbad.cds.unistra.fr/simbad/}}, and labelled them with H1$\sim$H9 in Figure \ref{fig:rgb}(a). 
And their physical properties are listed in Table \ref{tab:HII}. 
Of these \textsc{Hii} regions, H2 is an evolved \textsc{Hii} region, which has expanded to be the N59-North bubble, while the rest are ultra-compact (UC) \textsc{Hii} regions. 
\cite{Paulson2024} identified six O/B-type stars in N59-North bubble region (refer to region R1 in their work) based on the flux of radio continuum emission. 

Figure \ref{fig:rgb}(b) zooms in the N59-North bubble region from panel (a), with its R, G and B colors encoded in $Herschel$ 250, 160, and 70 $\mu$m, respectively. 
In the figure, almost all O/B-type stars (except B2) are associated with a UC \textsc{Hii} region, indicating that these sites are very active in star formation. 
Interestingly, the molecular cloud near B2 shows a finger-like structure pointing towards the O-type star O1. This structure is typical of bright rim cloud (BRC), where the expanding \textsc{Hii} region illuminates the molecular cloud near B2. 
However, due to the preexisting dense core inside the cloud, the surrounding low-density molecular cloud is first blown away, leaving the dense core in place. 
Moreover, using the online tool $MaserDB$\footnote{\url{https://maserdb.net/}}, eleven 6.7 GHz methanol (CH$_3$OH) maser sources are detected in the N59 region, and five of which are associated with the UC \textsc{Hii} regions. As far as we know, methanol maser is a signature of the early stage of massive star formation \citep{Chen2020a, Chen2020b}, suggesting that the N59 bubble still has potential for massive star formation. 

\section{Observations}

\subsection{MWISP CO observations}

The observations of CO toward the N59 bubble are part of the Milky Way Imaging Scroll Painting (MWISP\footnote{\url{http://www.radioast.nsdc.cn/mwisp.php/}}) project (see \citealt{Su2019}) conducted by the PMO-13.7m single-dish millimeter telescope at Delingha in China. 
The antenna employs the 3$\times$3-beam Superconducting Spectroscopic Array Receiver (SSAR) system as its frontend \citep{Shan2012} to obtain three CO molecular lines, $^{12}$CO, $^{13}$CO, and C$^{18}$O ($J=1-0$) simultaneously under the on-the-fly (OTF) mode. 
And a total of 18 Fast Fourier Transform (FFT) spectrometers are used as its backend that provide a 1 GHz bandwidth with 16384 channels, yielding a frequency interval of 61 kHz and a velocity coverage of $\sim$ 2600 km s$^{-1}$. 
The channel separation and RMS noise level at a channel width are 0.158 km s$^{-1}$ and 0.48 K for $^{12}$CO data, and 0.167 km s$^{-1}$ and 0.3 K for $^{13}$CO and C$^{18}$O data. 
The telescope has a beam size of 55$^{\prime\prime}$ at 110 GHz ($^{13}$CO and C$^{18}$O) and 52$^{\prime\prime}$ at 115 GHz ($^{12}$CO), and a pointing accuracy of $\sim 5^{\prime\prime}$. 
The final data product was converted into three-dimensional (3D) FIST data cubes of each cell ($30^{\prime}\times30^{\prime}$) with a grid spacing of $30^{\prime\prime}$. 

In this paper, the data cubes are clipped in a field of $30^{\prime}\times30^{\prime}$ centred at ($l, b$) $\sim$ ($33^{\circ}.075$, $-0^{\circ}.05$) and a velocity range of $0\sim140$ km s$^{-1}$. 
Throughout this paper, the galactic coordinate system is utilized and the equinox is J2000.0, and velocities are all given with respect to the local standard of rest (LSR). 

\subsection{Archival data}
The near-infrared (NIR) J, H, K band photometric data from the UKIDSS-GPS data release \citep{Lawrence2007} and the mid-infrared (MIR) photometric data from the $Spitzer$ telescope \citep{Fazio2004} are used to select the disk-bearing young stellar object (YSO) candidates in the survey region. 
The MIR images at IRAC 3.6, 4.5, 5.8 and 8.0 $\mu$m and MIPS 24 $\mu$m from the $Spitzer$ telescope are use to trace the emission from warm dust. 
And the $Herschel$ images at 70, 160, 250, 350, and 500 $\mu$m from the Hi-GAL survey are adopted to trace the emission from cold dust. 
The 1.1 mm radio continuum emission from the Bolocam Galactic Plane Survey (BGPS, \citealt{Rosolowsky2010}) are used to select and study the dust clumps associated with the N59 bubble. 
The angular resolution of the 1.1 mm map is $\sim$33$^{\prime\prime}$. 
The \textsc{Hi} 21 cm radio continuum emission data (with an angular resolution of $\sim1^{\prime}$) extracted from the VLA Galactic Plane Survey (VGPS, \citealt{Stil2006}) and the high resolution ($\sim6^{\prime\prime}$) 20 cm radio continuum data extracted from the Multi-Array Galactic Plane Imaging Survey (MAGPIS, \citealt{Helfand2006}) are adopted to trace the ionized gas. 
The $^{13}$CO data retrieved from the Galactic Ring Survey (GRS) is used to investigate the large scale kinematic environment of N59. 
The survey was performed by the Boston University and the Five College Radio Astronomy Observatory (FCRAO), which maps the Galactic Ring in the $^{13}$CO ($J=1-0$) line with an angular and spectral resolution of $46^{\prime\prime}$ and 0.2 km s$^{-1}$, respectively (see \citealt{Jackson2006}). 
The 6.7 GHz methanol masers in the survey region were identified using the online tool, $MaserDB$, provided by \cite{Ladeyschikov2019}.

\section{Results}

\subsection{Molecular gas distributions}

\begin{figure}
    \includegraphics[width=0.94\linewidth]{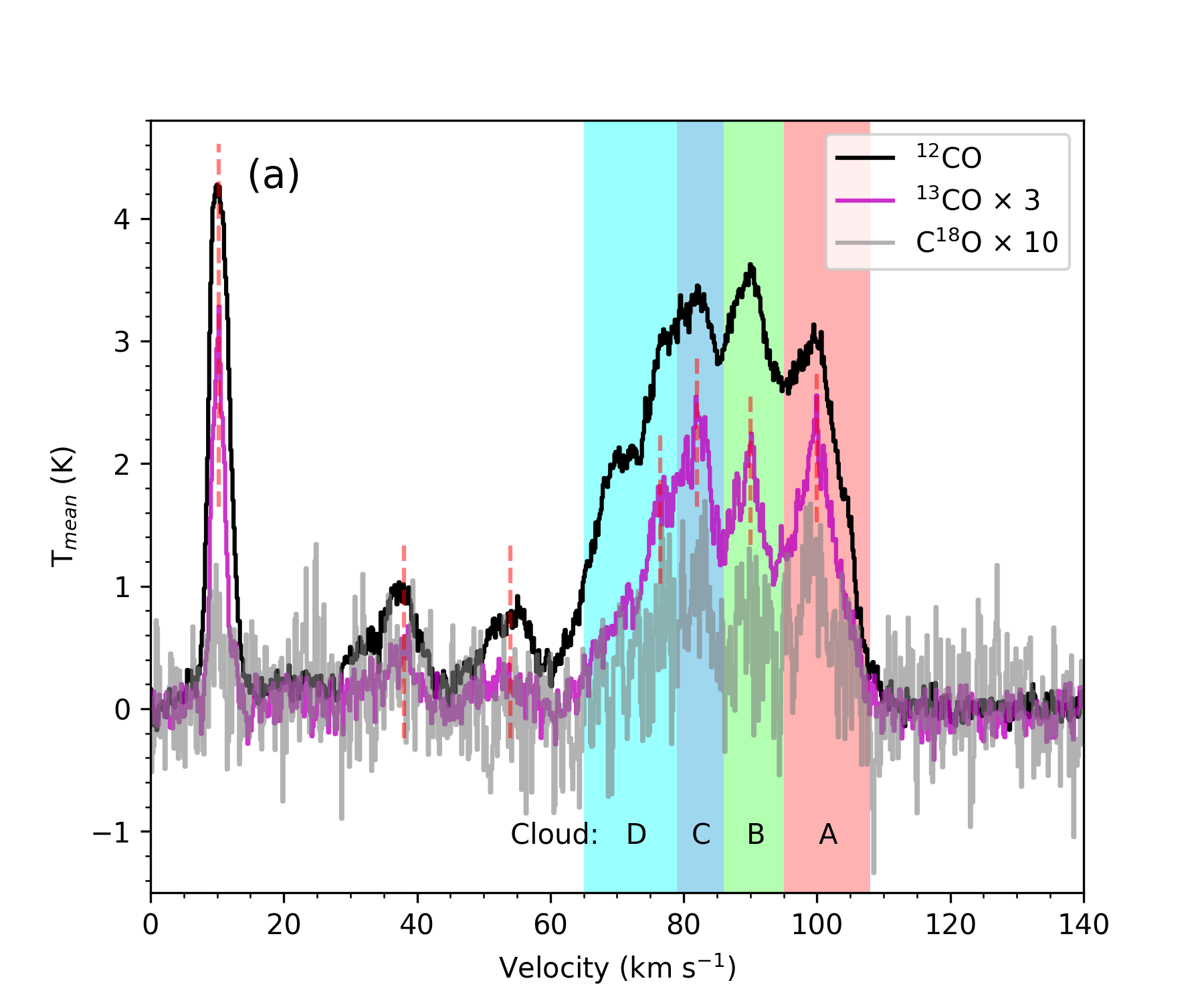}
    \includegraphics[width=0.94\linewidth]{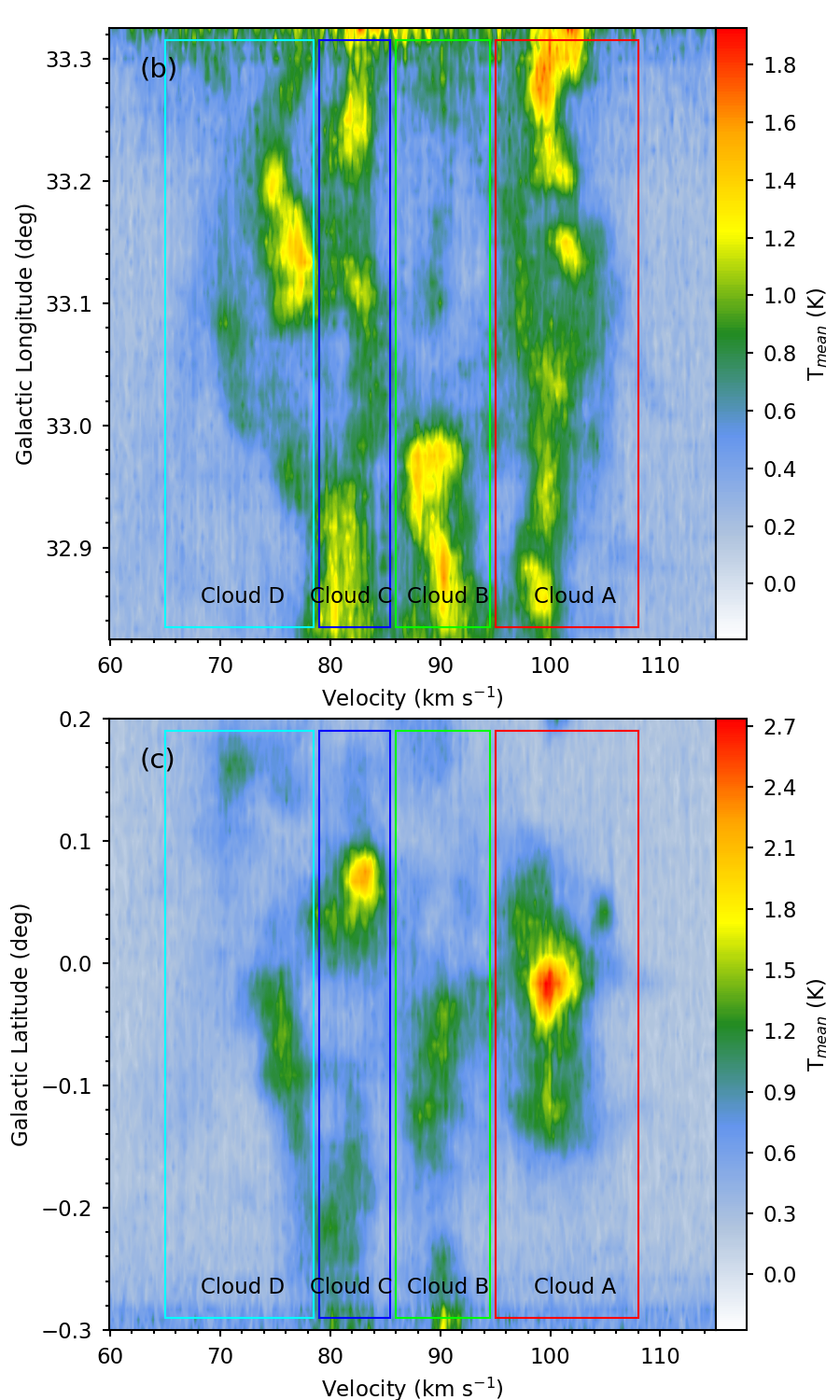} 
    \caption{Identification of molecular gas components. Panel (a) is the average spectra of the survey region. $^{12}$CO, $^{13}$CO and C$^{18}$O spectra are depicted as solid lines in black, magenta and gray, respectively. Filled squares in red, green, blue and cyan indicate the velocity windows of Cloud A, Cloud B, Cloud C and Cloud D, respectively. Panels (b) and (c) are the $l$-PV and $b$-PV diagrams of $^{13}$CO gas, respectively. Cloud A is in a velocity range of [95, 108] km s$^{-1}$. Cloud B is in a velocity range of [86, 95] km s$^{-1}$.
    Cloud C is in a velocity range of [79, 86] km s$^{-1}$.
    And Cloud D is in a velocity range of [65, 79] km s$^{-1}$. }
    \label{fig:pv}
\end{figure}

We extracted $^{12}$CO, $^{13}$CO and C$^{18}$O (J$=1-0$) line emissions simultaneously by the PMO-13.7m millimeter telescope. 
In general, $^{12}$CO usually traces extended molecular clouds due to its optical thick, while $^{13}$CO and C$^{18}$O usually trace dense molecular clouds due to their optical thin and low abundance. 
Figure \ref{fig:pv} (a) shows the average spectra of the $^{12}$CO, $^{13}$CO and C$^{18}$O ($J=1-0$) line emission in the survey region. 
The $^{12}$CO spectrum exhibits strong emission within the velocity range from 60 to 110 km s$^{-1}$, while the $^{13}$CO spectrum is mainly contained in a narrower velocity range of [65, 108] km s$^{-1}$. 
Within this wide velocity range, we identify four distinct spectrum peaks at 76 km s$^{-1}$, 82 km s$^{-1}$, 90 km s$^{-1}$ and 100 km s$^{-1}$. 
We depict these four components as Cloud D [65, 79] km s$^{-1}$, Cloud C [79, 86] km s$^{-1}$, Cloud B [86, 95] km s$^{-1}$ and Cloud A [95, 108] km s$^{-1}$, respectively. 
Note that outside of these velocity ranges, there are three significant peaks at 10 km s$^{-1}$, 38 km s$^{-1}$, and 54 km s$^{-1}$, which will not be considered in our study because they may belong to other spiral arms.

The $l$-PV and $b$-PV diagrams of the $^{13}$CO emission are shown in Figure \ref{fig:pv} (b) and (c). The four main components of Cloud A, Cloud B, Cloud C and Cloud D are highlighted by red, green, blue and cyan squares, respectively. 
These position-velocity (PV) diagrams show a significant coherent molecular cloud complex with a large velocity distribution from 65 to 108 km s$^{-1}$. 
According to the peak velocities of the four velocity components, the velocity separation is 10 km s$^{-1}$ along the light-of-sight between Cloud A and Cloud B, 18 km s$^{-1}$ between Cloud A and Cloud C, 6 km s$^{-1}$ between Cloud C and Cloud D, and 24 km s$^{-1}$ between Cloud A and Cloud D. 
The coherence (bridge feature) of these four components over such large velocity separations suggests the possibility of very complex cloud-cloud collision (CCC) processes between them, which will be discussed further in Section \ref{sec:CCC}.

\begin{figure*}
	\includegraphics[width=0.88\linewidth]{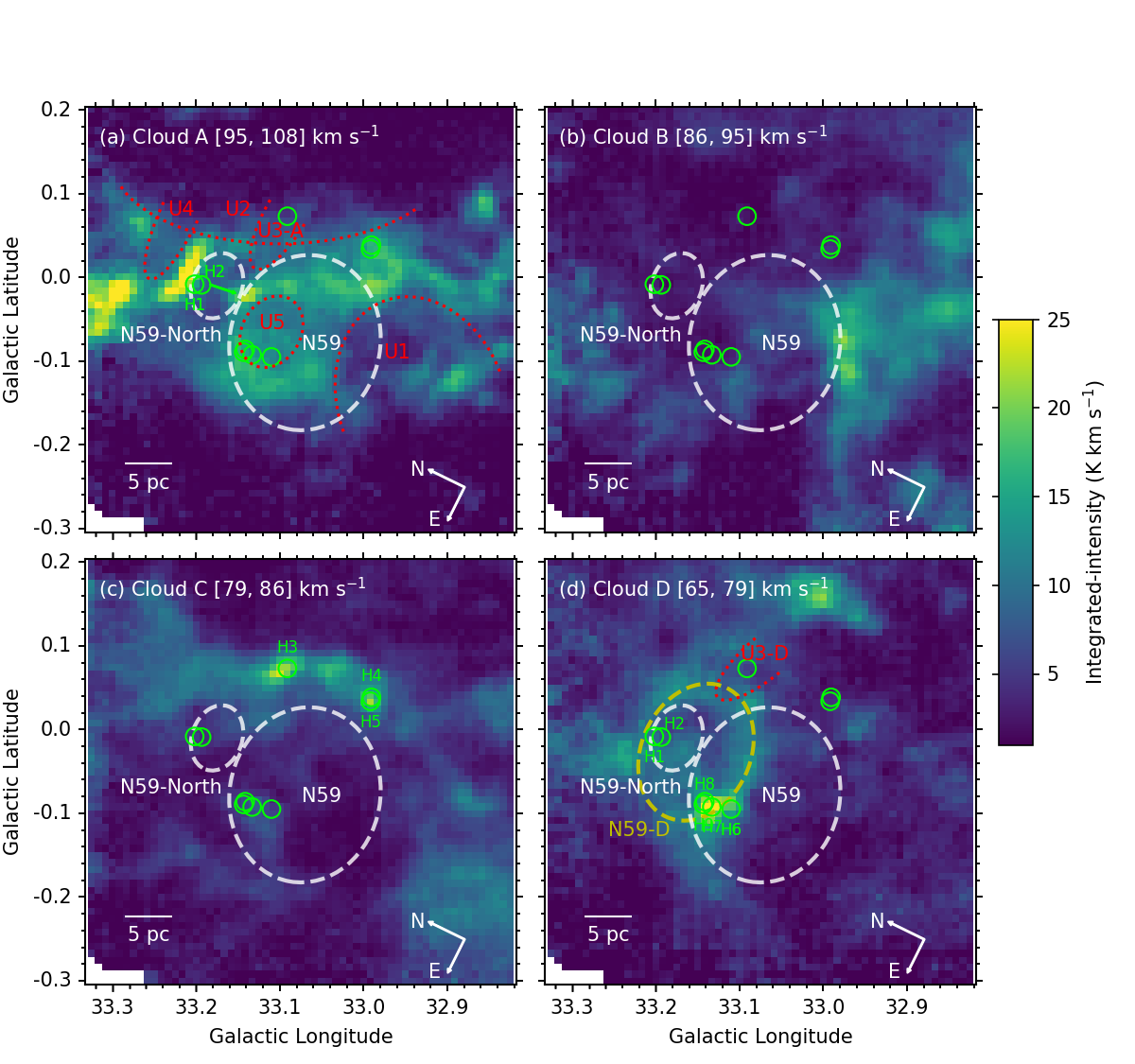}
    \caption{$^{13}$CO gas distribution of the four gas components. The N59 and N59-North bubbles are depicted by white dashed circles, and nine \textsc{Hii} regions are marked by green circles. The red dashed lines in panels (a) and (d) highlight the low density contours in Cloud A and Cloud D, the U-shape cavities defined in this paper. A bubble shape is shown in Cloud D which is depicted by a dashed yellow circle (namely the N59-D bubble). }
    \label{fig:comps}
\end{figure*}

Figure \ref{fig:comps} shows the $^{13}$CO gas distribution of these four gas components. 
The gas of Cloud A (see Figure \ref{fig:comps}(a)) is relatively smoothly distributed over the study area, with a significant small ring structure in the north. 
We named this ring structure the N59-North bubble, which was created by the expanding \textsc{Hii} region of H2. 
The "North" here refers to the northern direction of the Equatorial coordinate system, which, together with the "East" direction, are depicted in the lower right corner in each panel of Figure \ref{fig:comps}. 
And we hereafter use this coordinate system and the center of the N59 bubble to describe relative orientation between molecular clouds. 
An interesting bright rim cloud (BRC) can be seen along the southern direction of the massive star radiation in H2. 
Based on the lower gas density distribution (contours) of Cloud A, we found five distinct cavity structures, and labelled them with U1, U2, U3-A, U4 and U5 in Cloud A. 
The cavities exhibit a U-shape structure, suggesting that they may be the cavities created by CCC as described in the review of \cite{Fukui2021}. 
The gas of Cloud B (see Figure \ref{fig:comps}(b)) is distributed in a North American-like structure to the south edge of the N59 bubble, with no apparent \textsc{Hii} region counterparts. 
The gas of Cloud C (see Figure \ref{fig:comps}(c)) is distributed in an arc-like shell to the west of the N59 bubble, which is embedded with three UC \textsc{Hii} regions (H3, H4 and H5), indicating active massive star formation in Cloud C. 
The gas in Cloud D (see Figure \ref{fig:comps}(d)) is basically distributed in a ring-like structure, which is named the N59-D bubble. 
The physical properties of each gas component are briefly listed in Table \ref{tab:vel_range}.

\begin{table*}
    \centering
    \caption{Physical properties of $^{13}$CO velocity components. }
    \begin{tabular}{lcclccccl}
    \hline
       Cloud  &  $l$  &  $b$  &  $v_{range}$ & $v_{peak}$ & R$_{eff}$ & Mass & <N$_{\rm H_2}$> & Associated \\
       name   &  ($^{\circ}$) & ($^{\circ}$) & (km s$^{-1}$)  & (km s$^{-1}$) & (pc) & (10$^4$ M$_{\odot}$) & ($10^{21}$ cm$^{-2}$)  & \textsc{Hii} regions \\
    \hline
       Cloud A  & 33.096 & -0.031  & $95\sim108$  &  100 & 10.6 & 8.7 & 4.7  & H1, H2\\
       Cloud B  & 32.951 & -0.073 & $86\sim95$    &  90  & 9.0  & 7.0 & 2.8  & ...\\
       Cloud C  & 33.039 & -0.015 & $79\sim86$    &  82  & 14.4 & 6.5 & 2.4  & H3, H4, H5\\
       Cloud D  & 33.130 & 0.006  & $65\sim79$    &  76  & 9.3  & 5.8 & 3.2  & H6, H7, H8, H9\\
    \hline
    \end{tabular}
    \label{tab:vel_range}
\end{table*}

From the view of line-of-sight projection, the N59-North bubble is surrounded by the N59-D bubble, which have similar shapes and position angles but in different sizes (twice as large), indicating that the N59-D bubble may be a gas envelope of the N59-North bubble. 
Since the N59-North bubble is shaped by the expanding \textsc{Hii} region (H2) in Cloud A, it is possible that the N59-D bubble in Cloud D is also shaped by the same \textsc{Hii} region. This indicates that Cloud A and Cloud D are physically in contact with each other, located at the same place in space, despite the large relative velocity of 24 km s$^{-1}$ between them. This is the most direct evidence of the CCC process. 
Notably, the southern edge of the N59-D bubble contains a high dense gas clump embedded with four UC \textsc{Hii} regions (H6$\sim$H9), suggesting that the clump is very active in star formation. 
Interestingly, the presence of a small cavity (named the U3-D cavity) to the west of N59-D is essentially the same size and shape as the U3-A cavity in Cloud A, suggesting that they may be cross-sections of the same cavity in different clouds. 
This is another evidence of physical contact between Cloud D and Cloud A, which will be discussed in Section \ref{sec:CCC}. 
We also fit the six U-shape cavities (U1, U2, U3-A, U3-D, U4 and U5) found in the study region by elliptic arcs, and list their physical properties in Table \ref{tab:U-shape}.

\subsection{Signatures of cloud-cloud collisions}\label{sec:CCC}

\begin{figure*}
    \includegraphics[width=0.98\linewidth]{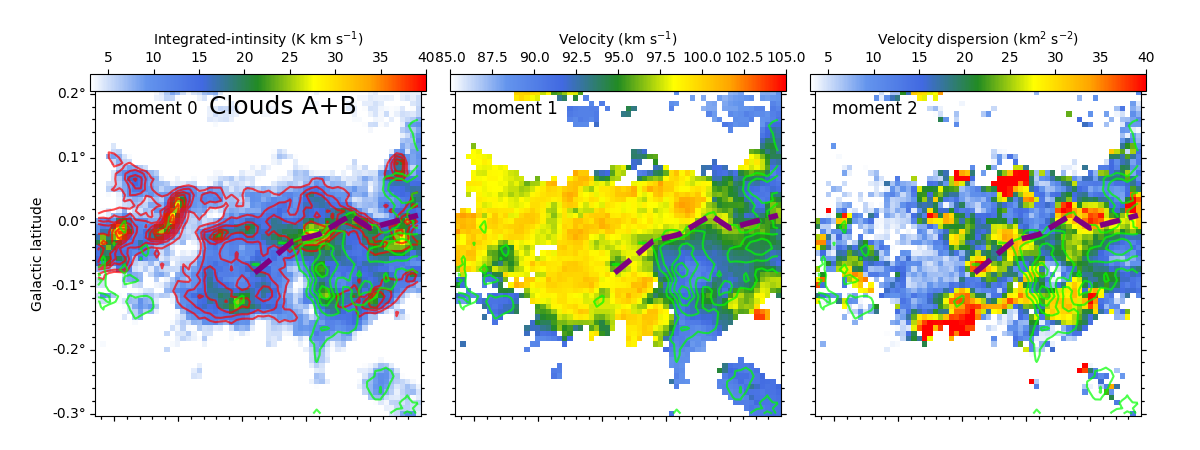} 
	\includegraphics[width=0.98\linewidth]{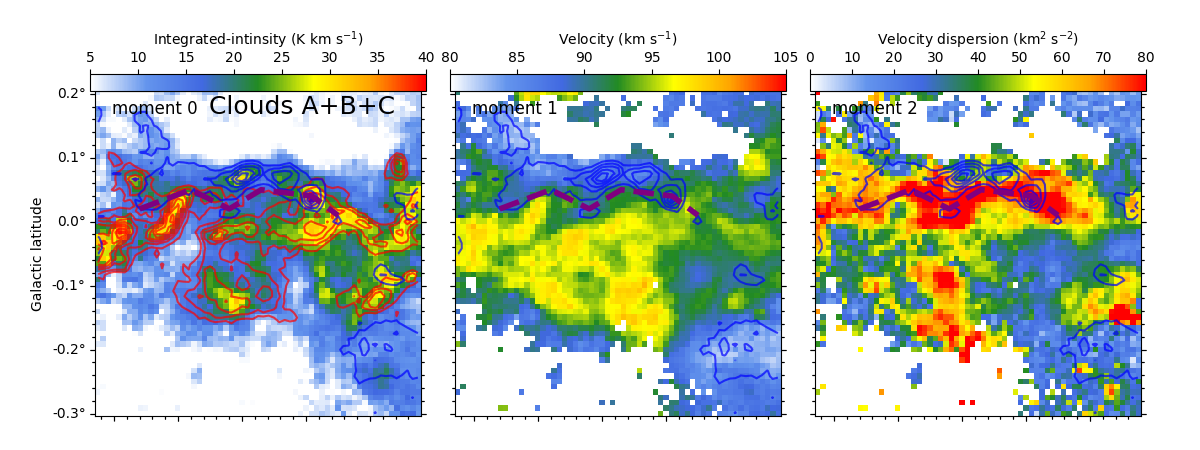}
    \includegraphics[width=0.98\linewidth]{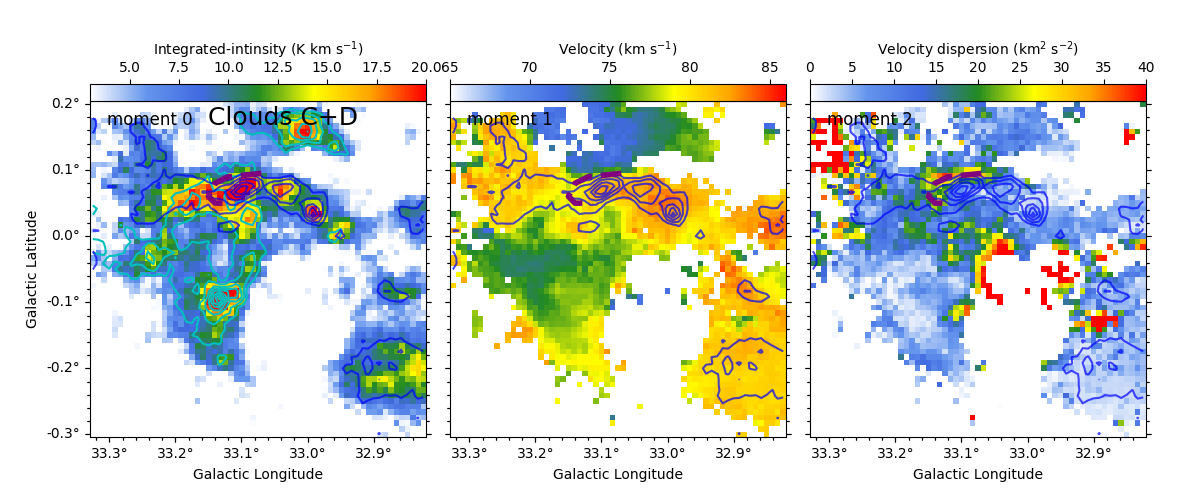}
    \caption{$^{13}$CO momentum maps of different gas components. The top row shows the three momentum maps (moment 0, moment 1 and moment 2, respectively) of Cloud A+B in a velocity range from 86 to 108 km s$^{-1}$. The middle row shows the three momentum maps of Cloud A+B+C in a velocity range from 79 to 108 km s$^{-1}$. The bottom row shows the three momentum maps of Cloud C+D in a velocity range from 65 to 86 km s$^{-1}$. Large velocity dispersion (> 30 km$^{2}$ s$^{-2})$ is highlighted by a purple dashed line in the moment 2 map, which indicates the possible colliding interface of two components. The same interface is also drawn in the moment 0 and moment 1 maps as well. Cloud A, Cloud B, Cloud C and Cloud D are drawn in Red, Green, Blue and Cyan contours respectively, which start from 9 to 30 K km s$^{-1}$ in steps of 3 K km s$^{-1}$. }
    \label{fig:moments}
\end{figure*}

We extract the possible colliding interfaces of CCCs based on three momentum maps: the integrated-intensity map (moment 0), the intensity-weighted mean velocity map (moment 1) and the velocity dispersion map (moment 2) of a spectral cube data. 
Figure \ref{fig:moments} shows the $^{13}$CO  momentum maps of different gas components. 
In the figure, the top, middle and bottom rows are the three momentum maps (moment 0, moment 1 and moment 2, respectively) of Cloud A+B in a velocity range of $86\sim108$ km s$^{-1}$, Cloud A+B+C in a velocity range of $79\sim108$ km s$^{-1}$ and Cloud C+D in a velocity range of $65\sim86$ km s$^{-1}$, respectively. 
Large velocity dispersion (> 30 km$^{2}$ s$^{-2})$ is highlighted by a purple dashed line in the moment 2 map, which probably indicate the colliding interface of two components or their overlapping along line-of-sight. 
Moreover, the interface is also consistent with the boundary of the two velocity components in the moment 1 map, indicating that the large velocity dispersion is mainly coursed by the collision of two clouds rather than a simple overlap of them. 
As shown in Figure \ref{fig:moments}, we detect three possible colliding interfaces, namely interface A-B, interface A-C and interface C-D.

\begin{figure*}
    \includegraphics[width=0.9\linewidth]{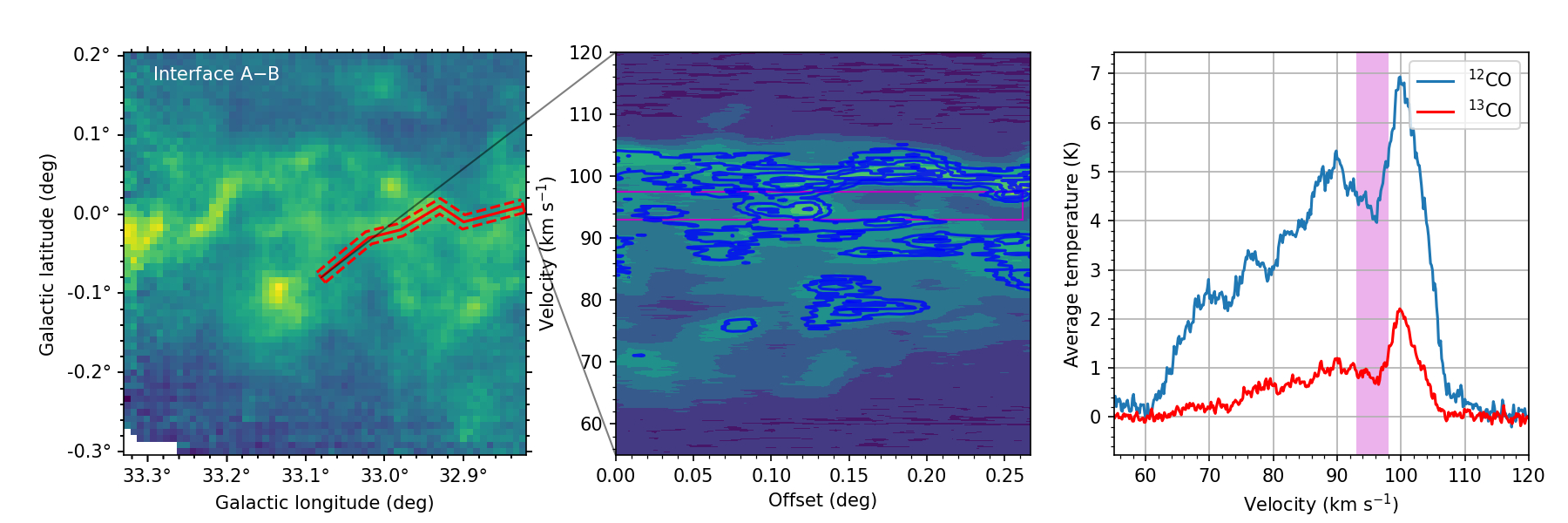} 
	\includegraphics[width=0.9\linewidth]{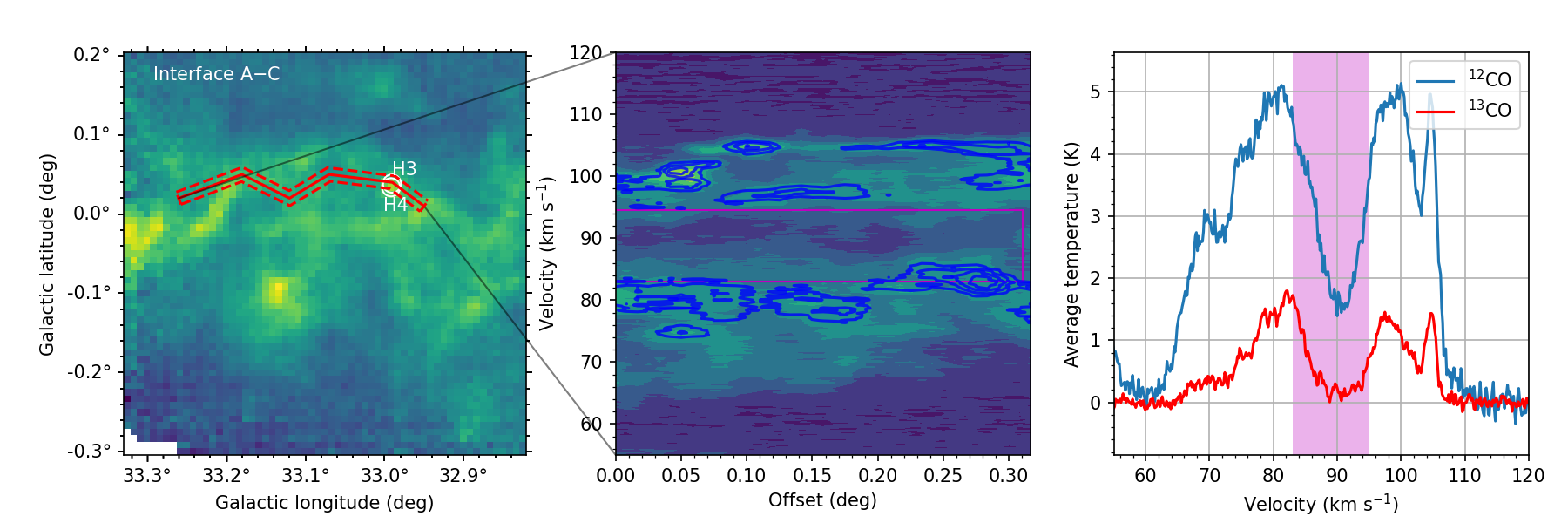}
    \includegraphics[width=0.9\linewidth]{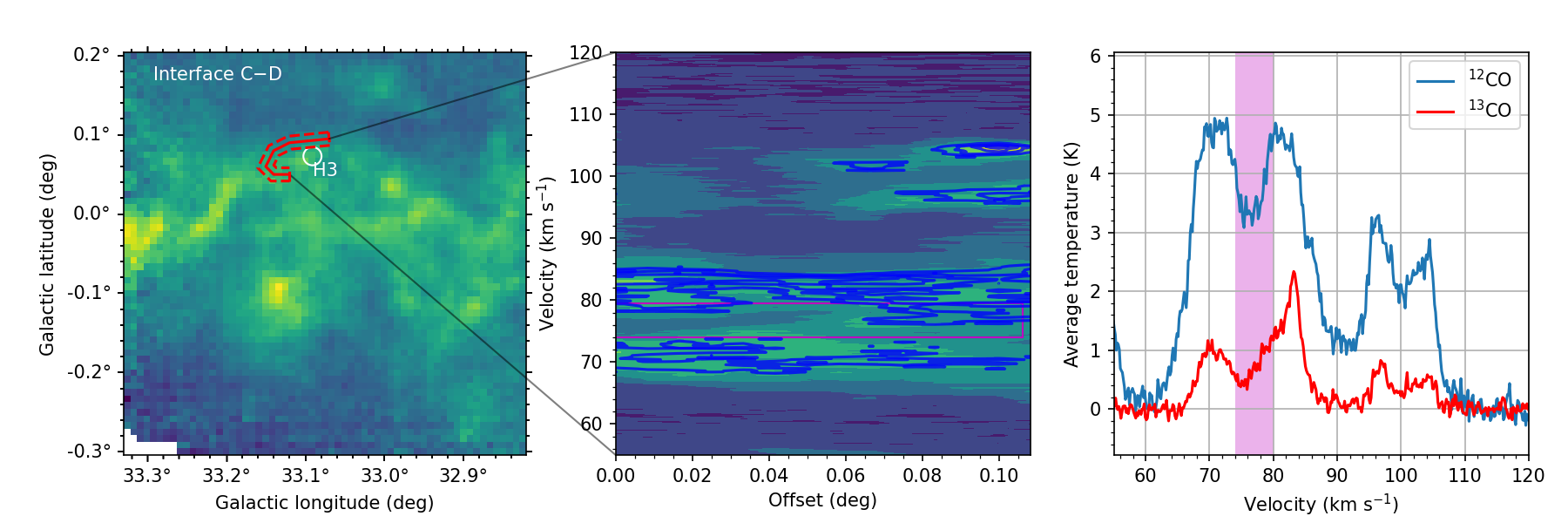}
    \includegraphics[width=0.9\linewidth]{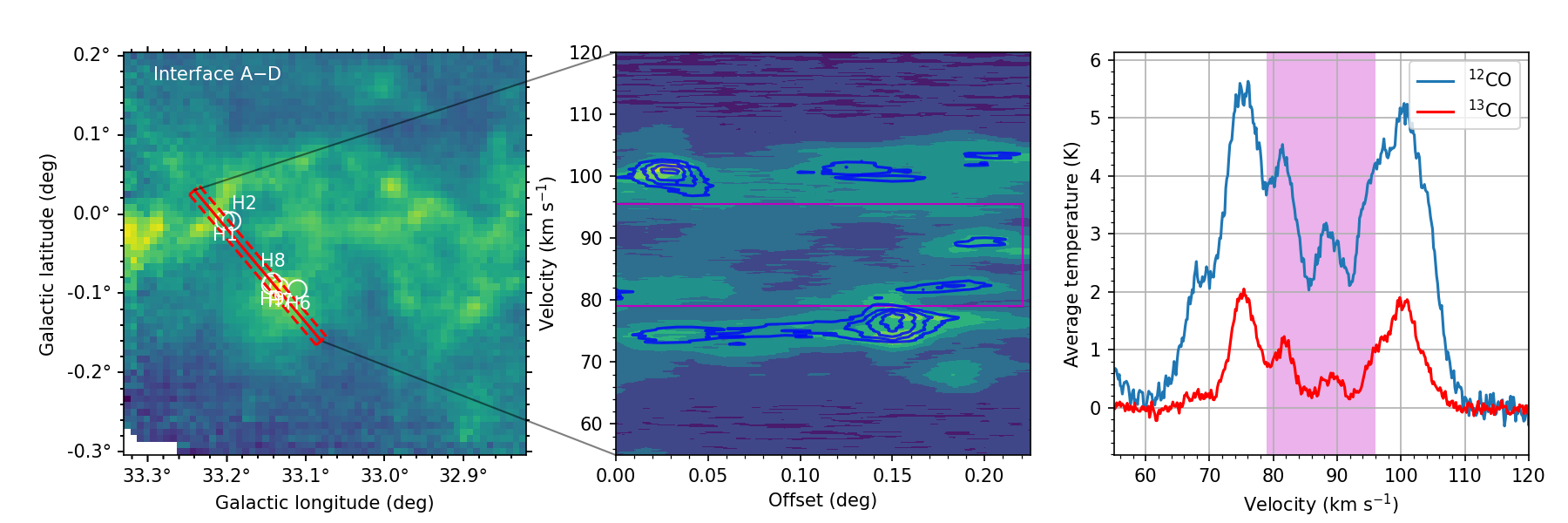}
    \caption{PV slices along colliding interfaces. The PV slices of interface A-B, interface A-C, interface C-D and interface A-D are shown in the first, second, third and fourth row, respectively. To all rows, the left panel shows the path of PV cut overlaying on the $^{13}$CO integrated intensity map. The middle panel shows the PV slice of the PV cut, in which the background image is $^{12}$CO PV map while the blue contours are $^{13}$CO PV map. The right panel shows the average spectra of the PV slice. The bridge features are highlighted by magenta squares in the middle (open) and right (filled) panels. }
    \label{fig:inter}
\end{figure*}

We further investigate broad bridge feature by extracting the position-velocity (PV) slices along the colliding interfaces. 
Figure \ref{fig:inter} shows the PV slices along the colliding interfaces of A-B, A-C, C-D and A-D that are presented in Figure \ref{fig:moments}. 
Note that interface A-D is simply the straight line between the brightest emission peaks of Cloud A and Cloud D, because they overlap each other in the line-of-sight direction. 
The PV slices (middle panels) and their average spectra (right panels) reveal broad bridge features at the intermediate velocity range of two components. 
As shown in the figure, broad bridge features are detected in a velocity range of [93, 98] km s$^{-1}$ for interface A-B, [86, 95] km s$^{-1}$ for interface A-C, [76, 81] km s$^{-1}$ for interface C-D and [80, 96] km s$^{-1}$ for interface A-D. 
The properties of these colliding interfaces are listed in Table \ref{tab:interface}.

\begin{table}
    \centering
    \caption{Properties of colliding interfaces. }
    \begin{tabular}{cccll}
    \hline
       CCC  & $v_{col}$ &   $v_{bridge}$  &  Triggered &  Triggered \\
       pairs  &  (km s$^{-1}$)  &  (km s$^{-1}$)  & \textsc{Hii} regions & YSO groups \\
    \hline
       A-B        & 10       & $93\sim98$  & ...  & C6, C8$\sim$C12\\
       A-C        & 18       & $83\sim95$  & H3, H4, H5  & C2, C3, C4, C5\\
       C-D        & 6        & $74\sim80$  & H3  & C3\\
       A-D        & 24       & $79\sim96$  & H1, H2, H6$\sim$H9  & C1, C13\\
    \hline
    \end{tabular}
    \label{tab:interface}
\end{table}

\begin{figure*}
	\includegraphics[width=\linewidth]{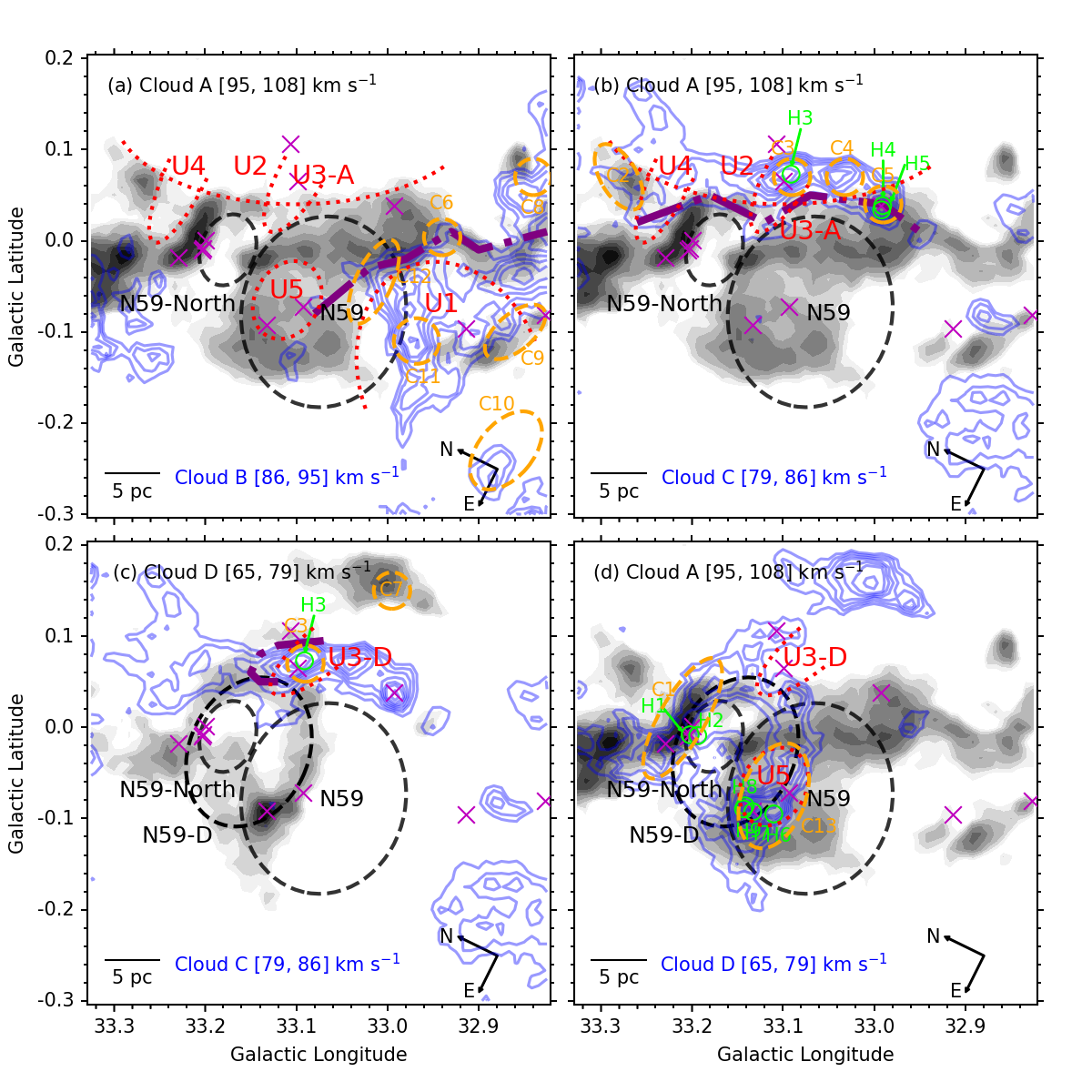}
    \caption{Complementary distributions of colliding clouds. Panel (a) Complementary distribution of Cloud A and Cloud B. The gray scale filled contours indicate Cloud A, while the blue contours indicate Cloud B. Panel (b) Complementary distribution of Cloud A and Cloud C. The gray scale filled contours indicate Cloud A, while the blue contours indicate Cloud C. Panel (c) Complementary distribution of Cloud C and Cloud D. The gray scale filled contours indicate Cloud D, while the blue contours indicate Cloud C. Panel (d) Complementary distribution of Cloud A and Cloud D. The gray scale filled contours indicate Cloud A, while the blue contours indicate Cloud D. To all panels, the contour levels are drawn in $(0.25, 0.3, 0.35, 0.4, 0.45, 0.5, 0.6, 0.7, 0.8, 0.9, 0.95)\times 32$ K km s$^{-1}$, and the purple dash-dotted lines indicate the colliding interfaces. The magenta crosses indicate the 6.7 GHz masers, and the green circles indicate the \textsc{Hii} regions. The red dashed lines indicate the U-shape cavities as shown in Figure \ref{fig:comps}. And the orange dashed ellipses indicate the fitting of YSO groups as shown in Figure \ref{fig:YSOs_dist}. }
    \label{fig:complementary}
\end{figure*}

Many studies (i.e., \citealt{Torii2015, Torii2017, Fukui2018}) have pointed out that the broad bridge feature is a significant evidence of CCC process, which represents the coherent gas slowing down during collision. 
Based on the three observational signatures of CCC summarised in \cite{Fukui2021} and mentioned in Section \ref{sec:intro}, we also investigate the spatial distribution and cavity structure among the four components to verify the existence of CCC. 
As shown in Figure \ref{fig:complementary}, a total of four possible complementary molecular pairs and six U-shape cavities are found. 
In Figure \ref{fig:complementary}(a), we detect five U-shape cavities, namely U1, U2, U3-A, U4 and U5 in Cloud A according to its low density distribution. A distinct complementary distribution can be seen in the southern part of Cloud A, where the U1 cavity is fully filled by Cloud B. 
This nearly perfect concave-convex structure is strong evidence for the existence of CCC. Meanwhile, the colliding interface A-B (purple dash-dotted line) from momentum analysis also outlines the junction of Cloud A and Cloud B, which reinforces the proof of CCC. 
In addition, the U1 cavity of Cloud A also coincides with the opening of the N59 bubble, indicating that the N59 bubble might be broken by the collision from Cloud B.

Another perfect complementary distribution is found between Cloud A and Cloud C (as shown in Figure \ref{fig:complementary}(b)). 
Cloud C as a whole has a slightly fluctuating arc-like structure, just completely fills the U2, U3-A and U4 cavities in Cloud A. 
The formation of this multiple cavity structure in large cloud (Cloud A) may be caused by the irregular distribution of Cloud C before collision. 
Two small protrusion structures in Cloud C were inserted into Cloud A by collision, resulting in the formation of the U3-A and U4 cavities. 
This premature collision increased the compression time of gas, leading to the formation of a dense core in Cloud C, which in turn formed a massive star that evolved into a UC \textsc{Hii} region (H3). 
Interestingly, the southern end of Cloud C intersects with Cloud A, suggesting an earlier collision there. This collision resulted in the formation of the densest core in Cloud C. Two embedded UC \textsc{Hii} regions (H4 and H5) are found in this dense core, suggesting that massive star formation there may have been triggered by CCC. 
The colliding interface A-C (purple dash-dotted line) outlines the boundary of Cloud C and fits the cavities in Cloud A simultaneously, suggesting a clear interaction between Cloud C and Cloud A.

The spatial distribution of Cloud D (see Figure \ref{fig:complementary}(c)) exhibits a ring-like structure (the N59-D bubble), with a small opening to the west that can be fitted by the U-shape cavity U3-D. 
It can be seen that the position and shape of the U3-D cavity are basically the same as that of the U3-A cavity in Cloud A, indicating that the U3-D and U3-A cavities may be the cross-sections of the same cavity (i.e., U3) in different components. 
Therefore, it is reasonable to infer that Cloud C may collide with Cloud A and Cloud D at the same time, leaving the U3-A and the U3-D cavities in Cloud A and Cloud D, respectively. 
Similar to the collision between Cloud C and Cloud A, the collision between Cloud C and Cloud D compresses Cloud C gas and triggers the formation of the massive star in the H3 UC \textsc{Hii} region. 

In Figure \ref{fig:complementary}(d), the spatial distribution of Cloud D overlaps with Cloud A in line-of-sight, with no obvious complementary distribution between them. 
However, the similarity of the ring-like structure (the N59-D bubble) of Cloud D to the N59-North bubble of Cloud A (twice as large) suggests a physical interaction between the two clouds. 
Stellar winds from the massive star in the evolved \textsc{Hii} region (H2) shaped both the N59-North bubble and the N59-D bubble, indicating that Cloud A and Cloud D are located in the same place, implying a CCC process. 
This high-speed ($\sim 24$ km s$^{-1}$) collision, as a consequence, triggered the formation of massive stars in the H1 and H2 \textsc{Hii} regions in Cloud A, as well as young massive stars in the H6, H7, H8 and H9 UC \textsc{Hii} regions in Cloud D.

\subsection{Young stellar population}

In order to characterize regions of ongoing star formation, it is always required to have a detailed knowledge of young stellar objects (YSOs) and their clustering behavior in a given star-forming region. 
\cite{Paulson2024} have investigated the YSOs in the N59 bubble using the near- and mid-infrared color-color (CC) and color-magnitude (CM) diagrams. 
According to their results, a total of 59 YSOs (28 Class I and 31 Class II YSOs) were found within a search radius of 9$^{\prime}$ centered at $(l, b)=(33^{\circ}.068, -0^{\circ}.044)$. 
They used color-color diagrams of the mid-infrared (MIR) IRAC I1-I2-I3-I4 bands and the near- (NIR) and mid-infrared H-K-I1-I2 bands (i.e., phase 1 and phase 2 in the method of \citealt{Gutermuth2009}) to select YSOs. 
However, the phase 2 of their scheme did not work well enough to separate the Class II YSOs from the photometric sources, hence their results abandoned a large number of Class II sources. 
Moreover, their scheme did not take into account sources in the longer MIR band (i.e., $Spitzer$-MIPS 24 um), which could reveal protostars that are more deeply embedded in molecular clouds. 
Based on this, we employed four different MIR/NIR schemes, including CM diagrams of [3.6]$-$[24]/[3.6] and H$-$K/H, to identify more embedded and reddened YSO candidates.
Before running the schemes, we firstly retrieved the $Spitzer$-IRAC four bands (I1 [3.6], I2 [4.5], I3 [5.8] and I4 [8.0]), $Spitzer$-MIPS 24 $\mu$m band, and UKIDSS J, H, K band point source catalogs in a $30^{\prime}\times30^{\prime}$ box region centered at $(l, b) = (33^{\circ}.075, -0^{\circ}.05)$. And then cross-matched these three catalogs within $1^{\prime\prime}$ radius to construct a total point source catalog. 
Detail criteria of the schemes are given in \cite{Dewangan2018} and \cite{Baug2018}, and the steps are as follows:

\begin{figure*}
	\includegraphics[width=\linewidth]{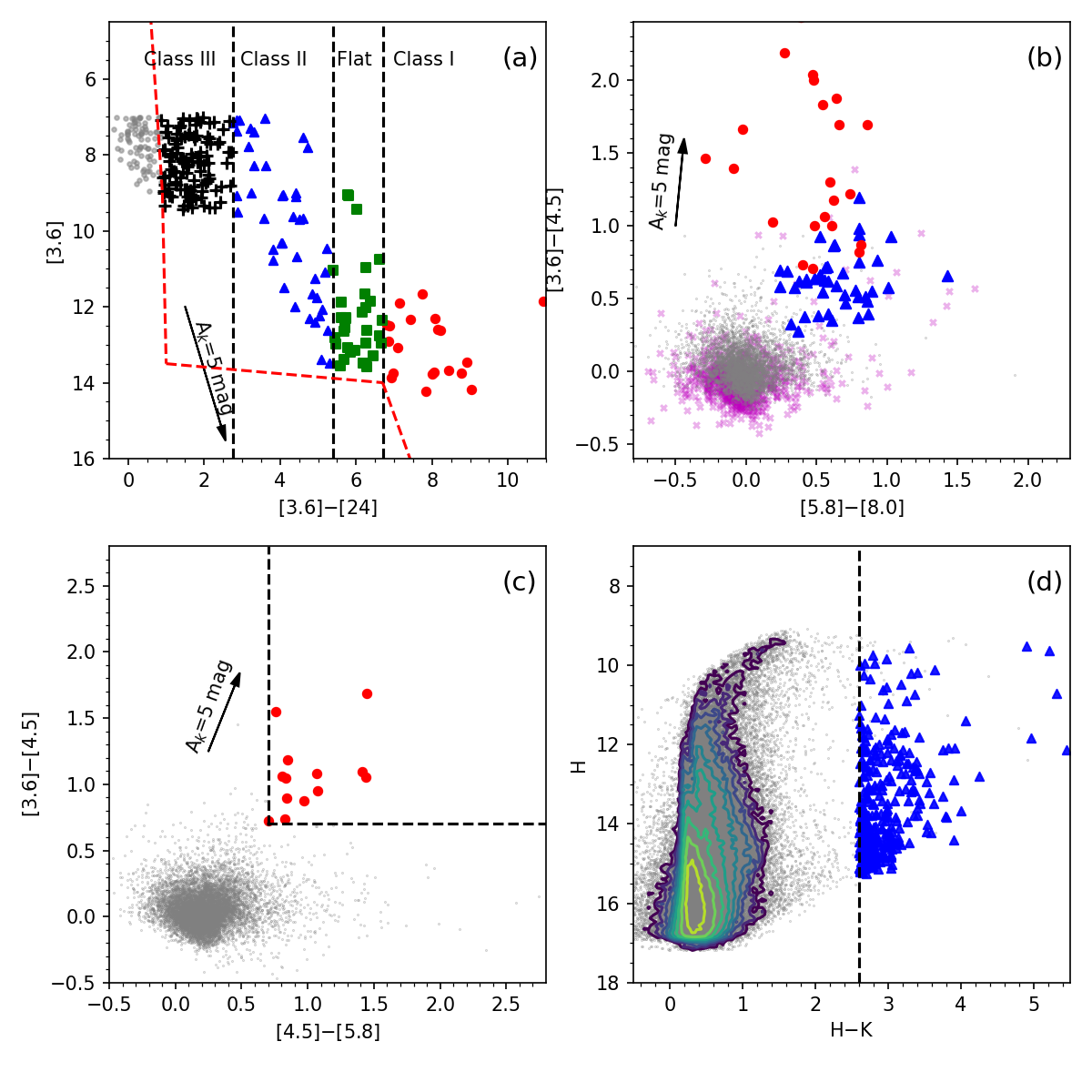}
    \caption{YSOs classification schemes using the MIR and NIR color-color (CC) and color-magnitude (CM) diagrams. (a) The [3.6]$-$[24]/[3.6] CM diagram. The red dashed line separates the YSOs from the field stars, and the black dashed lines classify them into Class I (red filled circles), Flat-spectrum (green filled squares), Class II (blue filled triangles) and Class III (black pluses), respectively. An extinction vector corresponding to A$_k=5$ mag is also shown. More detail criteria can be found in \protect\cite{Rebull2011}. (b) The [5.8]$-$[8.0]/[3.6]$-$[4.5] CC diagram of point-like sources detected in all four $Spitzer$-IRAC bands. The possible contaminants are marked with magenta crosses according to the scheme in \protect\cite{Gutermuth2009}. (c) The [4.5]$-$[5.8]/[3.6]$-$[4.5] CC diagram of point-like sources from the first three $Spitzer$-IRAC bands. The criteria are given in \protect\cite{Hartmann2005} and \protect\cite{Getman2007}. (d) The H$-$K/K CM diagram of those sources that are only detected in the H and K bands. The gray dots indicate the field stars from a nearby reference field ($30^{\prime}\times30^{\prime}$ box centered at $l=33^{\circ}.30$, $b=-0^{\circ}.55$). And the contours indicate the 2D histogram of number density of field stars. A cutoff value of H$-$K of 2.6 mag is estimated from the distribution of the field stars. }
    \label{fig:YSOs}
\end{figure*}

(i) [3.6]$-$[24]/[3.6] CM diagram: We firstly separated out young embedded sources using their MIR magnitudes at 3.6 and 24 $\mu$m. 
The $Spitzer$-IRAC point sources were retrieved from the GLIMPSE I catalog, and the $Spitzer$-MIPS 24 $\mu$m point-like sources were retrieved from the MIPSGAL catalog. For all IRAC and MIPS bands, we ensured that the errors of each band are less than 0.2 mag. 
Figure \ref{fig:YSOs}(a) shows the [3.6]$-$[24]/[3.6] color-magnitude diagram of these point sources, and the color criteria of YSO classification are given in \cite{Guieu2010} and \cite{Rebull2011}. 
Finally, we identified a total of 93 YSOs (20 Class I, 31 Flat-spectrum and 42 Class II) and 112 Class III sources following this $Spitzer$ CM scheme. 

(ii) IRAC I1-I2-I3-I4 CC diagram: Although MIPS 24 $\mu$m point sources are good at revealing protostars embedded in molecular clouds, they also suffer from strong nebulosity, resulting in few detection. 
Therefore, we constructed a series of color-color schemes of IRAC four bands to identify disk-bearing YSOs. 
The criteria are given in the Phase 1 method described in \cite{Gutermuth2009}. 
In this method, we eliminated contaminants (e.g., PAH/SF galaxies, AGN galaxies, shock-knots and PAH apertures) prior to classifying the YSOs. 
And then identified Class I YSOs based on the criteria in the [4.5]$-$[5.8]/[3.6]$-$[4.5] CC diagram. 
After excluding the above sources, the Class II YSOs were selected based on the criteria in the [4.5]$-$[8.0]/[3.6]$-$[5.8] CC diagram. All the classified sources of this scheme are shown in [5.8]$-$[8.0]/[3.6]$-$[4.5] CC diagram in Figure \ref{fig:YSOs}(b). 
Accordingly, a total of 66 YSOs (23 Class I and 43 Class II) and 942 contamination sources were identified using this scheme. 

(iii) IRAC I1-I2-I3 CC diagram: Since 8.0 $\mu$m still suffers from nebulosity, we additionally constructed the [3.6]$-$[4.5]/[4.5]$-$[5.8] CC diagram to identify YSOs (see Figure \ref{fig:YSOs}(c)). Simple criteria of [4.5]$-$[5.8]$\geq$0.7 and [3.6]$-$[4.5]$\geq$0.7 were employed to identify Class I YSOs \citep{Hartmann2005, Getman2007, Baug2018}. Using this scheme, a total of 13 Class I YSOs were identified. 

(iv) H$-$K/H CM diagram: Due to the higher sensitivity of UKIDSS in NIR bands, a large number of faint YSOs that are ignored by IRAC but visible in UKIDSS. 
Therefore, we further used H$-$K/K CM diagram to select faint, reddened YSOs. 
The J, H, K source table was retrieved from the UKIDSS-GPS tenth archival data release (UKIDSSDR10plus) catalog, which was recently published with complete and reliable photometric data. 
Due to the presence of circumstellar material, YSOs are redder than nearby field stars, therefore, a color H$-$K cutoff can be used to simply select the possible YSOs. 
The cutoff H$-$K$=$2.6 was estimated from the H$-$K/K CC diagram of a nearby referent field region (a $30^{\prime}\times30^{\prime}$ box centered at $(l,b)=(33^{\circ}.3, -0^{\circ}.55)$). Figure \ref{fig:YSOs}(d) illustrates the positions of field stars (gray dots and density contours) and the possible Class II YSOs (blue triangles) in H$-$K/K CM diagram. 
Using this scheme, a total of 342 Class II YSO candidates (red sources) were identified above the cutoff value. 

\begin{figure*}
	\includegraphics[width=\linewidth]{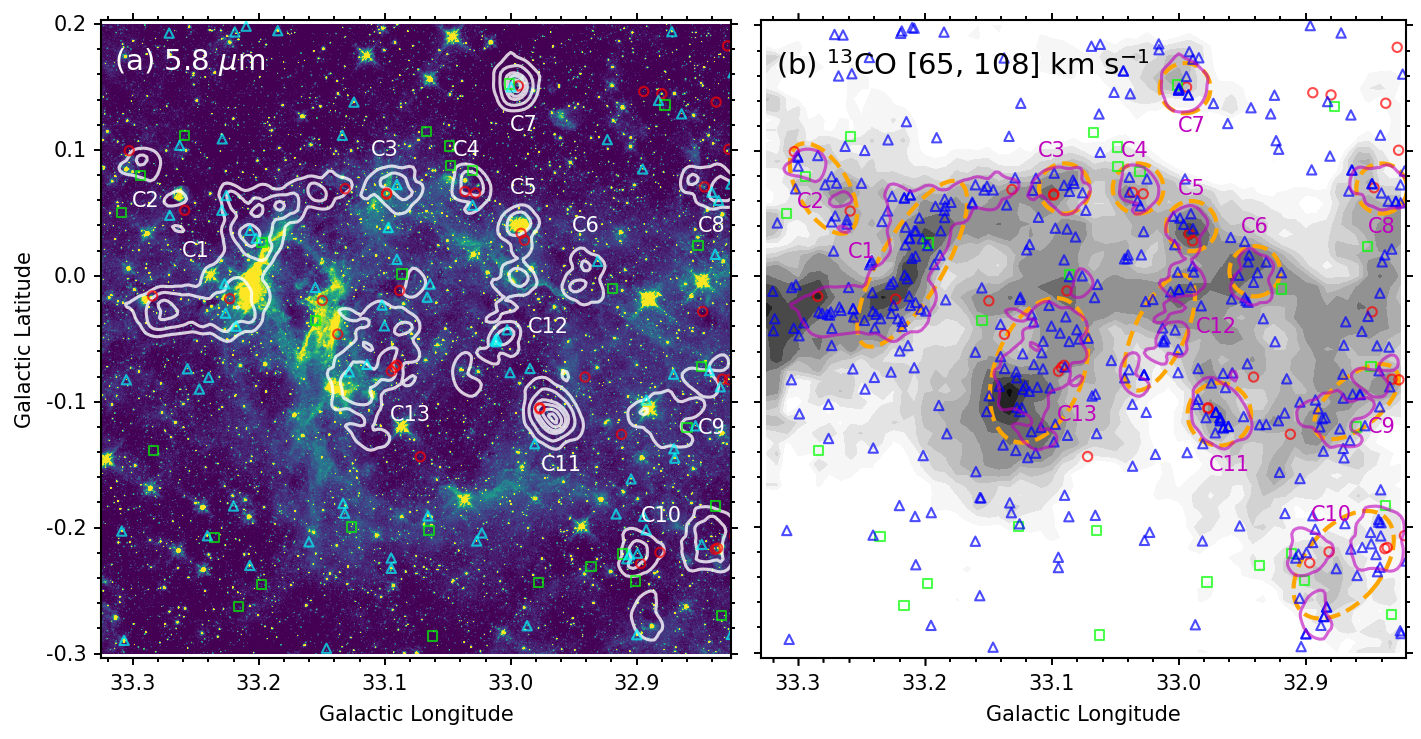}
    \caption{Spatial distribution of YSOs in the N59 region. (a) Spatial distribution of YSOs overlays on the 5.8 $\mu$m image. Only YSOs selected from $Spitzer$ MIR bands are shown. Class I, Flat-spectrum and Class II YSOs are marked by red circles, green squares and skyblue triangles, respectively. The white contours are the 6 nearest-neighbors surface density of all the YSOs (including the red sources from UKIDSS), drawn at 0.6, 1, 2, 3, 5, and 7 YSOs pc$^{-2}$. YSOs are mainly clustered into 13 YSO groups marked by C1$-$C13. (b) Spatial distribution of all the YSOs overlays on the $^{13}$CO integrated-intensity map for an integrated velocity range of $65-108$ km s$^{-1}$. Class I, Flat-spectrum and Class II YSOs are marked by red circles, green squares and blue triangles, respectively. The magenta contours outline the 6 nearest-neighbors surface density at 0.6 YSOs pc$^{-2}$. }
    \label{fig:YSOs_dist}
\end{figure*}

As a result, our YSO classification schemes yield a total of 514 YSOs (56 Class I, 31 Flat-spectrum and 427 Class II), and 112 Class III in the $30^{\prime}\times30^{\prime}$ survey region. 
All the identified Class I, Flat-spectrum and Class II YSOs are distributed over the $^{13}$CO integrated-intensity map in Figure \ref{fig:YSOs_dist}(b). Remarkably, a large number of YSOs are clustered in the ridges of molecular clouds.

In order to learn more about the clustering of YSOs, we perform the surface density distribution toward all the selected 514 YSOs. 
We compute the surface density map base on a given number of the nearest-neighbour (NN) YSOs divided by its enclosed circle area. 
More details of the NN technique can be found in \cite{Gutermuth2009}, \cite{Bressert2010} and \cite{Dewangan2018}. 
Taking into account the FOV size ($30^{\prime}\times30^{\prime}$) and distance (4.66 kpc), we construct a $10^{\prime\prime}$ grid and a number of 6 NN to calculate the surface density. 
Figure \ref{fig:YSOs_dist}(a) shows the surface density contours of YSOs overlaid on the $Sptizer$-IRAC 5.8$\mu$m image. The YSOs density contour levels are 0.6, 1, 2, 3, 5, 7 YSOs pc$^{-2}$. 
And we also highlight the boundaries of surface density by drawing the contour only at 0.6 YSO pc$^{-2}$ in Figure \ref{fig:YSOs_dist}(b). 
As shown in the figure, the Class I YSOs mainly embed in the most dense part of the molecular cloud, while the Flat-spectrum and Class II YSOs are slightly scattered around the ridges of the molecular cloud. 
For all YSOs, they trend to cluster into at least 13 individual subgroups (marked by C1$-$C13), which generally coincide with the position of the molecular clumps. 
We also use ellipses to fit these YSO groups to obtain their physical properties, and list them in Table \ref{tab:YSO}. As a result, a total of 300 YSOs ($\sim60\%$) are clustered in groups. 
Due to the complexity of N59 velocity components, the association between these YSO groups and each component, as well as their relationship with CCCs (see Figure \ref{fig:complementary}), will be discussed in Section \ref{sec:sf}.

\setlength{\tabcolsep}{3pt}
\begin{table}
    \centering
    \caption{Properties of YSO groups fitted by ellipses. }
    \begin{threeparttable}
    \begin{tabular}{cccccccccc}
    \hline
       YSO  & $l$ &   $b$  &  a &  b & PA & N$_{\rm H_2, peak}$\tnote{a} & N$_{\rm YSO}$\tnote{b} & N$_{\rm MS}$\tnote{c} \\
       gruop & ($^{\circ}$) & ($^{\circ}$)  & ($^{\circ}$) & ($^{\circ}$) & ($^{\circ}$) & ($\times10^{21}$ cm$^{-2}$) & & \\
    \hline
       C1   &  33.210 & 0.010 & 0.07 & 0.02 & 60 & 19.8 & 79 & 5\\
       C2   & 33.280 & 0.070 & 0.04 &  0.02 & 300 & 8.7 & 15 & 0\\
       C3   & 33.090 & 0.070 & 0.02 & 0.02 & 0 & 12.3 & 10 & 2\\
       C4   & 33.032 & 0.070 & 0.02 &  0.02 & 0 & 8.8 & 9 & 0\\
       C5   & 32.990  & 0.040 & 0.02 & 0.02 & 0 & 10.0 & 8 & 3\\
       C6   & 32.940 & 0.000 & 0.02 & 0.02 & 0 & 4.3 & 12 & 0\\
       C7   & 32.995 & 0.150 & 0.02 & 0.02 & 0 & 11.7 & 10 & 0\\
       C8   & 32.840  & 0.070 & 0.02 & 0.02 & 0 & 7.0 & 10 & 0\\
       C9   & 32.860 & -0.100  & 0.04 & 0.02 & 40 & 6.0 & 22 & 0\\
       C10  & 32.870 & -0.230 & 0.05  & 0.03 & 50 & 6.8 & 32 & 0\\
       C11  & 32.968 & -0.110 & 0.03 & 0.03 & 0 & 10.1 & 17 & 0\\
       C12  & 33.015 & -0.045 & 0.05 & 0.02 & 65 & 7.1 & 20 & 0\\
       C13  & 33.110 & -0.075 & 0.06 & 0.04 & 70 & 18.0 & 55 & 6\\
    \hline
    \end{tabular}
    \begin{tablenotes}
        \scriptsize
        \item[a] The peak H$_2$ column density of each YSO group. 
        \item[b] The number of YSOs contained in each YSO group. 
        \item[c] The number of massive stars (O/B-type stars and masers) contained in each YSO group.  
    \end{tablenotes}
    \end{threeparttable}
    \label{tab:YSO}
\end{table}

\section{Discussion}

\subsection{Star formation triggered by cloud-cloud collisions}\label{sec:sf}

Recent studies \citep{Fukui2021, Chen2024} have shown that CCC can trigger star formation across a wide range of masses. 
Figure \ref{fig:complementary} shows the spatial distribution between the collision pairs and star formation tracers (YSO groups, \textsc{Hii} regions and masers). 
As shown in Figure \ref{fig:complementary}(a), five YSO groups (C6, C8, C9, C11 and C12) and two 6.7 GHz masers are associated with the spatially common sections of Cloud A and Cloud B. 
Therefore, it is more likely that these star formation activities were triggered by CCC. 
The detection of two masers suggests that the CCC has the potential to trigger massive star formation, although no associated UC \textsc{Hii} region has been found.

In Figure \ref{fig:complementary}(b), the YSO groups of C2, C3, C4, and C5 are distributed along the colliding interface. The UC \textsc{Hii} regions (H3, H4 and H5) and 6.7 GHz masers at C3 and C5 indicate massive star formation there. 
Since the protrusion structure of Cloud C at C3 leads to the formation of the U3-A cavity, the formation of the C3 YSO group and the massive star in the H3 UC \textsc{Hii} region may be triggered by CCC. 
In addition, C5, as the junction between Cloud A and Cloud C, is the site where the collision has the greatest impact, hence, the C5 YSO group, and massive stars in H4 and H5 UC \textsc{Hii} regions are more likely to be triggered by CCC. 
In contrast, the more distant C2 and C4 do not detect the formation of massive stars. 

As mentioned in Section \ref{sec:CCC}, Cloud C may collide with Cloud A and Cloud D at the same time, leaving the U3-A and U3-D cavities in Cloud A and Cloud D, respectively, so that the gas clumps within the U3 cavity may be compressed simultaneously by these two CCC processes. As shown in Figure \ref{fig:complementary}(c), the formation of the C3 YSO group and the massive star in the H3 UC \textsc{Hii} region may also be affected by the CCC between Cloud C and Cloud D in addition to the CCC between Cloud C and Cloud A. 

Figure \ref{fig:complementary}(d) shows that the distribution of the C1 YSO group is almost consistent with the northern dense clump of the N59-North bubble in Cloud A and the northern gas shell of the N59-D bubble in Cloud D, indicating that it may be triggered by the collision between Cloud A and Cloud D. However, due to the presence of the evolved \textsc{Hii} region (H2), it cannot be ruled out the triggering effect of the expanding \textsc{Hii} region. 
Moreover, the C13 YSO group is almost distributed in the U5 cavity, indicating that it may be triggered by the CCC of Cloud A and Cloud D. 
In particular, there is a dense clump embedded with four O/B-type stars at the inner edge of this cavity, therefore, these massive stars may be triggered by the collision of the tail of Cloud D with Cloud A.

\begin{figure}
	\includegraphics[width=\linewidth]{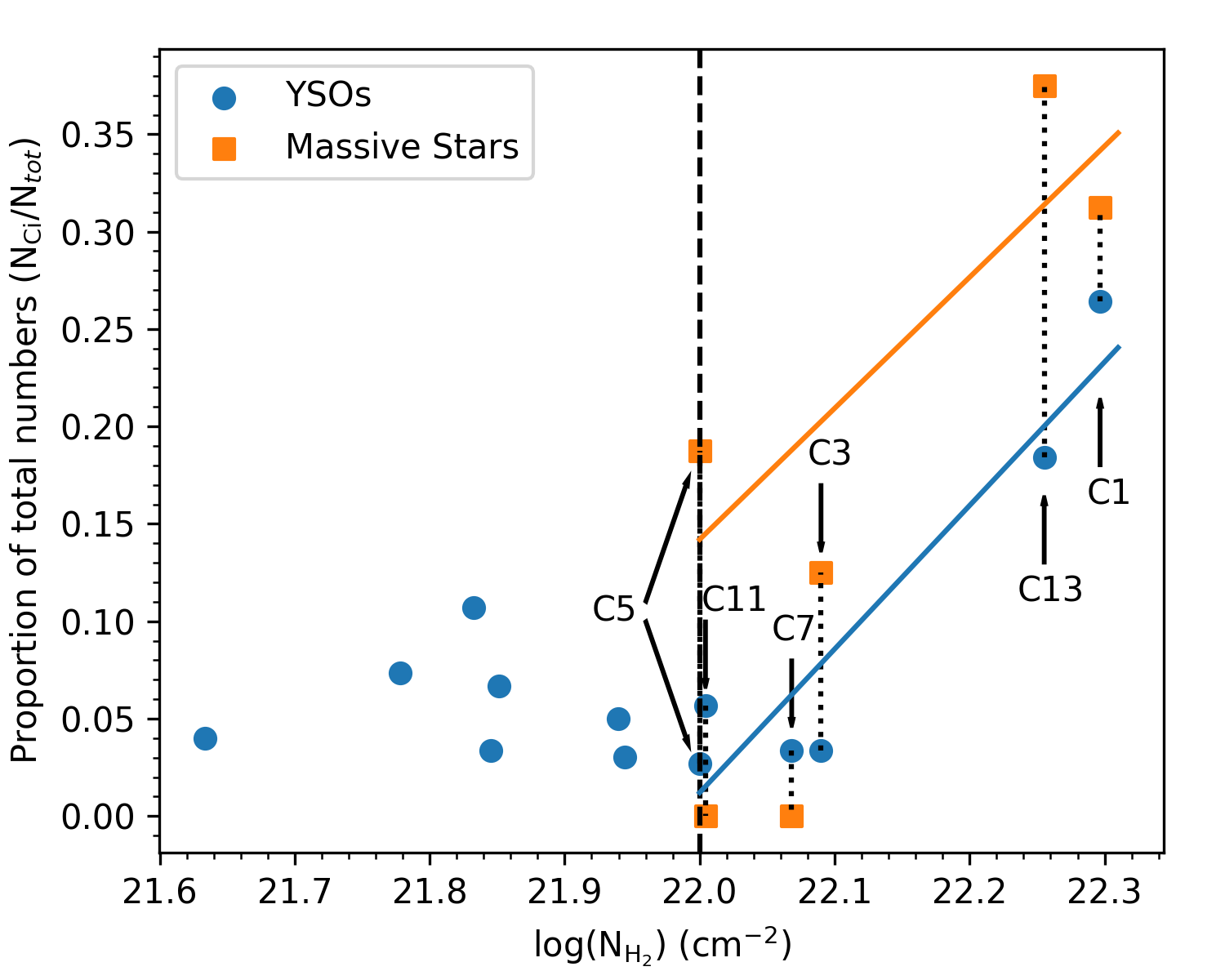}
    \caption{The relationship between the peak column density of molecular clouds within each YSO group and the proportion of the number of YSOs and massive stars it contains. 
    The blue dots and orange squares represent the number and proportion of YSOs and massive stars, respectively. 
    The blue and orange straight lines indicate the best-fitting lines (for data with N$_{\rm H_2}\geq10^{22}$ cm$^{-2}$ and Prop $>0$) in terms of YSOs and massive stars, respectively, with slopes of 0.74 and 0.67. 
    The dotted lines between the blue and orange data points indicate that the YSOs and massive stars belong to the same YSO group. 
    Six YSO groups (C1, C3, C5, C7, C11, and C13) with peak column densities greater than $10^{22}$ cm$^{-2}$ are marked by arrows. 
    The vertical dashed line indicates the threshold column density of $10^{22}$ cm$^{-2}$. }
    \label{fig:pNH2}
\end{figure}

The above discussion only studies the triggering relationship between CCC and star formation from the spatial distribution of the YSO groups and the colliding interfaces. 
We also investigate this relationship further by testing the relationship between the peak column density of the gas clumps within the YSO groups and the number of YSOs and massive stars formed by that clumps. 
The relationship is illustrated in Figure \ref{fig:pNH2}, where the Y-axis values are the proportions of the number of YSOs/Massive stars contained in each YSO group. 
In the Figure, the number of YSOs and massive stars increases with the peak column density when the peak column density is greater than 10$^{22}$ cm$^{-2}$ (the threshold column density). This is compatible with the conclusion of \cite{Fukui2021}. The plot exhibits a linear trend, which can be fitted with straight lines. Their best-fitting slopes are basically the same, about 0.7, which is consistent with the slope of $0.73\pm0.11$ reported in \cite{Enokiya2021}. These fits may not be accurate due to the lack of data points, but the upward trend is clear. Combined with Figure \ref{fig:complementary}, the massive stars are mainly formed in clumps within C1, C3, C5 and C13, which are located at the colliding interfaces (dominated by CCCs). In contrast, C7 and C11, far from the colliding interfaces (dominated by turbulence), do not have massive star formation, although their column densities exceed the threshold. This implies that at the same column density, CCC is more likely to trigger the formation of massive stars than turbulent only molecular clouds.

It is important to note that besides the CCC mechanism, other processes, such as the collect and collapse (C\&C) mechanism and compression of pre-existing clumps by the shock front around an expanding \textsc{Hii} region, could also trigger star formation. 
Hence, the star formation in N59 may be the result of a combination of multiple star formation processes. 
Star formation in a region is not coeval. According to the work of \cite{Evans2009}, the age of Class I YSOs is about $0.1\sim0.44$ Myr, while the age of Class II YSOs is about $1\sim2$ Myr. 
Since the duration of a typical CCC process is about 0.5 Myr,  younger Class I YSOs at the colliding interface may be triggered by CCC, while older Class II YSOs may be triggered earlier by other processes. As mentioned above, all the massive stars in N59 (except the massive star of B2) are associated with the CCC processes. Therefore, the age of a massive star (or the age of its ionized \textsc{Hii} region) can be used to estimate the lower limit of the duration of CCC process. 

Once a massive star forms, it emits a large number of ultraviolet (UV) photons, which ionize its surrounding hydrogen gas and then evolve into a \textsc{Hii} region. 
We estimate the age of the UC \textsc{Hii} region base on the radius of the initial Str$\ddot{\rm o}$mgren sphere \citep{Stromgren1939, Dyson1980}, 
\begin{equation}
    R_s = \Big(\frac{3\ \dot{N}_{LyC}}{4\pi\beta_*n_0^2}\Big)^{1/3} \approx 0.74\ \dot{N}_{49}^{1/3}n_3^{-2/3}\ \rm pc, 
\end{equation}
where $\beta_*\sim2\times10^{-13}$ cm$^{3}$ s$^{-1}$ is the recombination coefficient for atomic hydrogen. $n_3\equiv [n_0/10^3\ \rm cm^{-3}]$ is the initial number density of gas in a unit of 10$^{3}$ cm$^{-3}$, and $\dot{N}_{49}\equiv [\dot{N}_{LyC}/10^{49}\ \rm s^{-1}]$ is the hydrogen-ionizing photon rate in a unit of $10^{49} \rm s^{-1}$. Using the formula (2) from \cite{Condon1992}, we can estimate $\dot{N}_{LyC}$ from integrated flux density ($S_{\nu}$) of radio continuum emission as flollows, 
\begin{equation}
    \dot{N}_{LyC} \approx 7.54\times10^{46}\ \Big(\frac{T_e}{10^4 \rm K}\Big)^{-0.45}\Big(\frac{\nu}{\rm GHz}\Big)^{0.1}\Big(\frac{S_{\nu}}{\rm Jy}\Big)\Big(\frac{d}{\rm kpc}\Big)^2\ \rm s^{-1}, 
\end{equation}
where $T_e$ is the electron temperature, $\nu$ is the central frequency of the radio continuum emission, $S_{\nu}$ is the integrated flux density, and $d$ is the distance of the \textsc{Hii} region. 
Since massive star is originally formed from a dense core, the UC \textsc{Hii} region is also embedded in this dense core, with a typical number density of $n_0=10^{4-5}$ cm$^{-3}$. 
Considering that the physical size of the UC \textsc{Hii} region is close to the radius of the Str$\ddot{\rm o}$mgren sphere. Using the formula (4) from \cite{Whitworth1994}, we estimate the age of the UC \textsc{Hii} region as follows, 
\begin{equation}
    t_s\approx 0.042\ \dot{N}_{49}^{1/3}n_3^{-2/3}\ \rm Myr, 
\end{equation}
Taking the typical values of UC \textsc{Hii} region ($n_0=10^4\ \rm cm^{-3}$ and $T_e=10^4$ K) and the distance of N59 ($d=4.66$ kpc) into account, we obtain the ages of the UC \textsc{Hii} regions in N59 and list them in Table \ref{tab:HII}. The UC \textsc{Hii} regions are very young, ranging in ages from 1 to 4 kyr. Notably, our age estimation has a factor of 2 uncertainty due to the assumption of initial density. 

On the other hand, for an evolved \textsc{Hii} region (i.e., H2), we estimate its age using a dynamical age from \cite{Spitzer1978}, assuming spherical expansion, 
\begin{equation}
    t_{dyn} \approx 0.057\ \Big(\frac{R_s}{\rm pc}\Big)\Big(\frac{v_i}{\rm 10\ km\ s^{-1}}\Big)^{-1}\left(\Big(\frac{R_{\textsc{Hii}}}{R_s}\Big)^{7/4}-1\right)\ \rm Myr, 
\end{equation}
where $R_s$ is the radius of the Str$\ddot{\rm o}$mgren sphere, $R_{\textsc{Hii}}$ is the radius of the evolved \textsc{Hii} region, and $v_i$ is the sound speed of the ionized gas, typically $v_i\approx10$ km s$^{-1}$. 
Considering the radius ($R_{\textsc{Hii}}=2.85$ pc) and the mean number density ($n_0=5\times10^3\ \rm cm^{-3}$) of H2, we obtain a dynamical age of 2 Myr for H2. Moreover, using the formula (5) from \cite{Whitworth1994}, we obtain the radius and time of the shell when fragmentation starts, i.e., $R_{frag}\approx2.26\ \rm pc$, $t_{frag}\approx1.36\ \rm Myr$. $t_{dyn} \geq t_{frag}$ indicates that the fragmentation process has occurred in the periphery of H2. Therefore, the contribution of star formation near H2 (such as the C1 YSO group) from the expansion of the \textsc{Hii} region cannot be ignored. 

These results are also supported by the findings of \cite{Paulson2024} for the R1 region within the N59 bubble. 
The R1 region exhibits a broken ring gas structure with several YSOs at the edges, suggesting the possible impact of O/B-type massive stars. 
R1 also includes a significant bright rim cloud (BRC) structure. 
The "head" of the BRC hosts a bright condensation and is oriented toward the massive star, indicating that the BRC originates from the influence of ionizing radiation from the massive star. 
This process is often referred to the radiation-driven implosion (RDI) process \citep{Sandford1982}. 
\cite{Paulson2024} also found two velocity components in R1, where p11 and p12 have a velocity of about 70-80 km s$^{-1}$ (corresponding to Cloud D in the paper), while nearby p13 has a peak velocity of about 100 km s$^{-1}$ (corresponding to Cloud A). 
The significant velocity difference exceeding 15 km s$^{-1}$ between the two molecular clouds rules out a picture of stars forming through cloud-cloud collapse, but a picture of a CCC is possible. 
Therefore, star formation in the R1 region consists of two major processes: the CCC process that triggers the formation of massive stars and the process that the massive stars trigger the formation of second-generation stars.

\subsection{Scenario of sequential cloud-cloud collisions}\label{sec:scen}

The present work provides a detailed investigation of CCC processes in the N59 bubble. 
In the previous sections, we revealed four velocity components, namely Cloud A, Cloud B, Cloud C and Cloud D, toward the N59 region, which are coherent in velocity space and exhibit distinct complementary boundaries between the molecular pairs, such as Clouds A-B, Clouds A-C and Clouds C-D. 
Although the complementary boundary of the molecular pair of Clouds A-D is not detected due to the projection effect, the similarity between the N59-D bubble in Cloud D and the N59-North bubble in Cloud A indicate that Cloud D and Cloud A coexist spatially. 
Therefore, the above signatures indicate that there are four major CCC processes between the four components: i.e., Clouds A-B, Clouds A-C, Clouds C-D, and Clouds A-D.

\setlength{\tabcolsep}{1.7pt}
\begin{table}
    \centering
    \caption{Physical properties of U-shape cavities fitted with elliptic arcs. }
    \begin{threeparttable}
    \begin{tabular}{lcccccccccc}
    \hline
       Uid  &  $l$  &  $b$  &  a & b & PA & [$\phi_1$, $\phi_2$]\tnote{a} & H\tnote{b} & CCC & $v$ & t\tnote{c} \\
            &  ($^{\circ}$) & ($^{\circ}$) & ($^{\circ}$)  & ($^{\circ}$) & ($^{\circ}$) & (rad) & (pc) & pairs & (km/s)  & (Myr) \\
    \hline
       U1  & 32.93 & -0.15  & 0.10  & 0.13  & -20 & $[0, \pi]$ & 10.6 & A-B  & 10 & 1.06\\
       U2  & 33.10 & 0.14   &  0.10 &  0.20 & 90  & $[\frac{7\pi}{10}, \frac{14\pi}{10}]$ & 5.7 & A-C  & 18 & 0.32\\
       U3-A  & 33.09 & 0.08 & 0.08  & 0.03  & -60  & $[-\frac{\pi}{2}, \frac{\pi}{2}]$ & 6.5 & A-C & 18  & 0.36\\
       U3-D  & 33.06 & 0.09  & 0.08 &  0.03 & -40  & $[-\frac{\pi}{2}, \frac{\pi}{2}]$  & 2.4 & C-D & 6  & 0.40\\
       U4   & 33.22 & 0.08 & 0.09   & 0.02  & -65 & $[-\frac{\pi}{2}, \frac{\pi}{2}]$ & 7.3 & A-C & 18 & 0.41 \\
       U5   & 33.11 & -0.06 & 0.05  & 0.04  & -60 & $[0, 2\pi]$ & 7.3 & A-D & 24 & 0.30 \\
    \hline
    \end{tabular}
    \begin{tablenotes}
        \scriptsize
        \item[a] [$\phi_1$, $\phi_2$] is the range of the elliptic arc. 
        \item[b] H is the arc height. The arc heights of U1, U3-A and U4 are the semi-major axis of the fitting ellipses. The arc heights of U2 and U3-D are the distances from the arc to the string. The arc height of U5 is the major axis of the ellipse. 
        \item[c] t is the duration of CCC. Assuming that all CCCs has the same incidence angle ($\theta$) of 45$^{\circ}$, then t$=$H/($v$ tan$\theta$)$=$H/$v$. 
    \end{tablenotes}
    \end{threeparttable}
    \label{tab:U-shape}
\end{table}

We calculate the duration of each CCC simply by dividing the travel distance by the collision speed. 
The travel distance can be obtained by fitting the size of the U-shape cavities caused by CCCs, while the collision speed can be estimated by the relative velocity of two molecular clouds in the line-of-sight direction. 
Table \ref{tab:U-shape} exhibits the physical properties of U-shape cavities fitted with elliptic arcs. 
Assuming that all CCCs has the same incidence angle ($\theta$) of 45$^{\circ}$ (related to the light-of-sight direction), the duration of each CCC is t$=$H/($v$ tan$\theta$)$=$H/$v$, where H is the arc height, and $v$ is the relative velocity of two molecular clouds. 
Combined with Table \ref{tab:HII} and Table \ref{tab:U-shape}, we obtain the duration of these four CCCs. 
For CCC pair A-B, its duration is about 1 Myr (1.06 Myr of U1). 
For CCC pair A-C, its duration is about 0.4 Myr (0.32 Myr of U2, 0.36 Myr of U3-A and 0.41 Myr of U4). 
For CCC pair C-D, its duration is about 0.4 Myr (0.40 Myr of U3-D). 
For CCC pair A-D, there is a two-step collision: for Cloud A vs. Cloud D-main, the collision occurred before 2 Myr (according to the dynamic age of the H2 \textsc{Hii} region); for Cloud-A vs. Cloud D-tail, the collision occurred before 0.3 Myr (0.30 Myr of U5). 
It can be seen that the duration of CCC pairs A-C and C-D is basically the same, which reflects on the other hand that the collision between Cloud C and Clouds A\&D is simultaneous.

\begin{figure}
	\includegraphics[width=\linewidth]{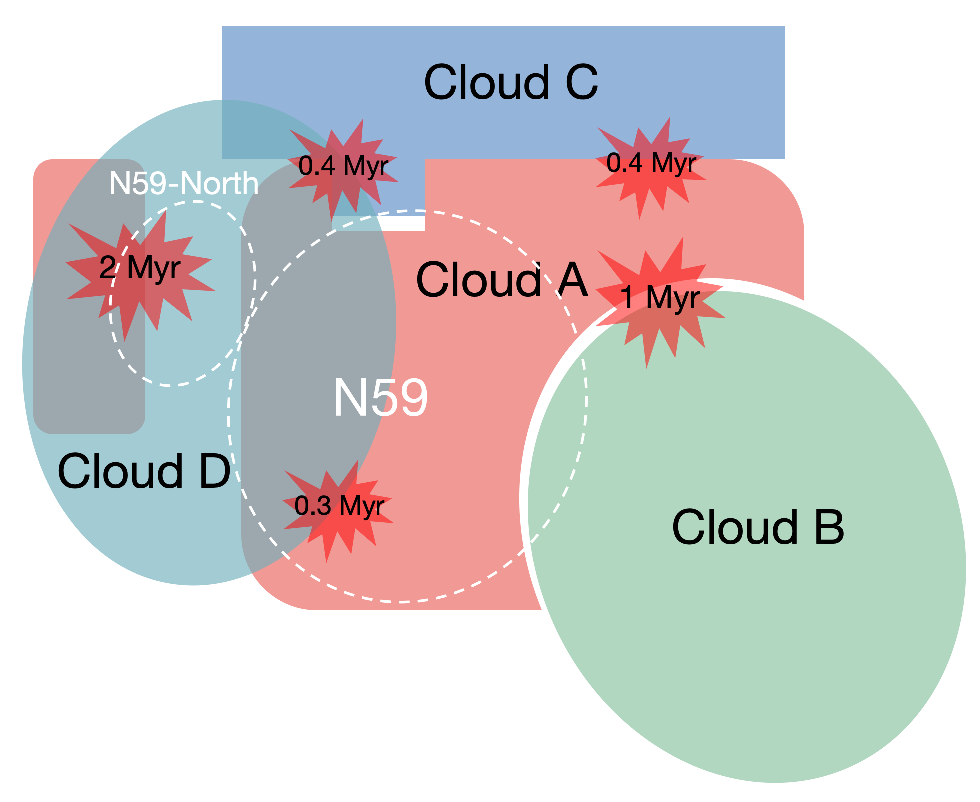}
    \caption{A simple scenario of sequential multiple cloud-cloud collisions in the N59 bubble. The time sequential order of collisions is: a) Cloud A collided with Cloud D about 2 Myr ago, b) Cloud B collided with Cloud A about 1 Myr ago, and c) Cloud C collided with both Cloud A and Cloud D about 0.4 Myr ago. }
    \label{fig:sketch}
\end{figure}

\cite{Habe1992} presented a simple scenario of CCC (hereafter the Habe model): a small cloud collides with a large cloud at supersonic speed, creating a U-shape cavity in the large cloud, compressing gas to form a dense layer, which in turn collapses to form stars in a wide range of mass. 
Based on this picture of CCC, we present a simplified sequential CCC scenario in N59 in Figure \ref{fig:sketch}. 

Firstly, Cloud A collided with Cloud D-main about 2 Myr ago, with a very large relative velocity of 24 km s$^{-1}$ (light-of-sight direction). 
The collision produced an O9.5V massive star, which further ionized its surrounding gas to form an expanding \textsc{Hii} region (H2). 
This process is consistent with the final phase of the Habe model, where the \textsc{Hii} region from the massive star erodes the colliding molecular clouds. 
However, the collision between Cloud D and Cloud A is not over yet. 
The collision also triggered the formation of four massive stars (O3, B5, B6 and B7) in the southern part of Cloud D (or the tail of Cloud D). 
This CCC started before 0.3 Myr, which is consistent with the ages (ranges from 1 to 4 kyr) of the UC \textsc{Hii} regions of these massive stars.

Secondly, Cloud B collided with Cloud A about 1 Myr ago, with a relative velocity of 10 km s$^{-1}$. 
This collision created the U1 cavity in Cloud A and triggered the formation of C6, C8, C9, C11, and C12 YSO groups, as well as two massive star formation candidates (6.7 GHz masers).

Finally, Cloud C collided with Cloud A and Cloud D simultaneously about 0.4 Myr ago, leaving two similar cavities in Cloud A (U3-A) and Cloud D (U3-D), respectively. 
Consequently, the collision triggered the formation of the C3 YSO group as well as the massive star in H3. 
Moreover, the collision of Cloud C with Cloud A also triggered the formation of the C5 YSO group and massive stars in H4 and H5. 
The ages of H3 (2 Kyr), H4 (1 Kyr), and H5 (3 kyr) as the lower limit timescale of CCC also support this collision.

As a result, the time sequential order of these multiple CCC processes is as follows: a) Cloud A collided with Cloud D about 2 Myr ago, b) Cloud B collided with Cloud A about 1 Myr ago, and c) Cloud C collided with both Clouds A and D about 0.4 Myr ago (see Figure \ref{fig:sketch}).

\subsection{Could cloud-cloud collision driven by SNR?}

In literature (\citealt{Kesteven1968, Zhou2011,Sofue2021}), there is a famous radio supernova remnant (SNR Kes 78, as shown in Figure \ref{fig:rgb}) stay close to the N59 bubble. 
\cite{Zhou2011} reported that Kes 78 is associated with the 81 km s$^{-1}$ molecular cloud (refer to Cloud C in this paper), but less evidence to be associated with the 91 km s$^{-1}$ molecular cloud (refer to Cloud B) and the 100 km s$^{-1}$ molecular cloud (refer to Cloud A). 
Their study revealed that the SNR Kes 78 evolved in an interstellar environment, with blast wave propagating velocity of $v_s= 1100$ km s$^{-1}$.
Combined with its radius of $r_s=17$ pc, the dynamical age of the remnant was estimated to be $t\sim2r_s/(5v_s)\sim6$ kyr according to the \cite{Sedov1959} evolution law. 
This suggests that the SNR Kes 78 is a very young SNR. If the CCC is driven by SNR, it can only be associated with the most recent CCC in N59. 
According to the above Section \ref{sec:scen} discussion, the most recent CCC occurred between Cloud C and Clouds A\&D. 
However, the duration of this CCC ($0.3\sim0.4$ Myr) is nearly 50 times longer than the age of the SNR ($\sim6$ kyr), suggesting that the collision has been occurred more earlier than the SNR shock-front hit Cloud C. 
Since the duration of this CCC is not comparable to the age of the SNR, it is impossible that the CCC is driven by the SNR.

\section{Conclusions}

The N59 bubble is a young and active massive star formation region. 
Our CO kinematic analysis, together with the YSO classification, reveals multiple cloud-cloud collision processes of the four components in the N59 bubble and their consequential impact on star formation. 
The important conclusions of this work are summarized as follows: 

\begin{enumerate}
    \item We propose that the N59 bubble is an active star formation region caused by multiple cloud-cloud collisions. 
    \item The gas in N59 can be decomposed into four velocity components, namely Cloud A [95, 108] km s$^{-1}$, Cloud B [86, 95] km s$^{-1}$, Cloud C [79, 86] km s$^{-1}$ and Cloud D [65, 79] km s$^{-1}$. 
    \item Four cloud-cloud collision processes occur involving the following velocity components: the 10 km s$^{-1}$ collision between Cloud A and Cloud B, the 18 km s$^{-1}$ collision between Cloud A and Cloud C, the 6 km s$^{-1}$ collision between Cloud C and Cloud D, and the 24 km s$^{-1}$ collision between Cloud A and Cloud D. 
    \item Using photometric data from $Spitzer$ and UKIDSS, a total of 514 YSO candidates are identified clustered in 13 YSO groups, and most of which ($\sim60\%$) are located along the colliding interfaces. 
    \item Cloud B collided with Cloud A, creating a distinct U-shape cavity of U1 in Cloud A and triggering the formation of at least five YSO groups, as well as two 6.7 GHz masers that trace the early stage of massive star formation. 
    \item Cloud C collided with Cloud A, creating a large U-shape cavity of U2 to the west of Cloud A and triggering the formation of at least three YSO groups and three massive stars. Additionally, due to the irregular boundaries of Cloud D, this CCC also creates two small U-shape cavities of U3-A and U4 in Cloud A, which may link with the formation of the C3 YSO group. 
    \item Cloud C collided with Cloud D, creating a small U-shape cavity of U3-D to the west of Cloud D. Since the position and shape of U3-D and U3-A are basically the same, they may be cross-sections of the same U-shape cavity in Cloud D and Cloud A. This indicates that Cloud C may collide with both Cloud A and Cloud D at the same time, which together trigger the formation of the C3 YSO group. 
    \item Cloud A collided with Cloud D-main, producing an O9.5V massive star with an expanding \textsc{Hii} region (H2). On the other hand, Cloud A collided with Cloud D-tail, creating a U-shape cavity of U5 in Cloud A and triggering the formation of four massive stars with UC \textsc{Hii} regions (H6, H7, H8 and H9). A large number of YSOs (C1 and C13 YSO groups) are also triggered by these two collisions as well as the expansion of \textsc{Hii} regions. 

    \vspace{0.5cm}
    Taken together, the star formation activity in the N59 bubble could be explained by sequential multiple cloud-cloud collision processes. 
    Firstly, Cloud A collided with Cloud D about 2 Myr ago, triggering the formation of an O9.5V massive star with an expanding \textsc{Hii} region (H2) and two YSO groups (C1 and C13). 
    Secondly, Cloud B collided with Cloud A about 1 Myr ago, triggering the formation of five YSO groups (C6, C8, C9, C11 and C12). 
    Finally, Cloud C collided with Cloud A and Cloud D simultaneously about 0.4 Myr ago, triggering the formation of three YSO groups (C3, C4 and C5) and three massive stars with UC \textsc{Hii} regions (H3, H4, and H5). 
\end{enumerate}

\section*{Acknowledgements}

We acknowledge supports by the National Key R\&D program of China (2022YFA1603102), the National Natural Science Foundation of China (NSFC, grant No. 11873002, 12011530065, 11590781). X.C. thanks to Guangdong Province Universities and Colleges Pearl River Scholar Funded Scheme (2019). 
This work made use of the data from the Milky Way Imaging Scroll Painting (MWISP) project conducted by Purple Mountain Observatory (PMO). MWISP was sponsored by National Key R\&D Program of China with grants 2023YFA1608000 \& 2017YFA0402701 and by CAS Key Research Program of Frontier Sciences with grant QYZDJ-SSW-SLH047. 
This paper made use of the NIR data products from the UKIRT Infrared Deep Sky Survey (UKIDSS) and the archived MIR data obtained with the $Spitzer$ Space Telescope (operated by the Jet Propulsion Laboratory, California Institute of Technology under a contract with NASA). 
The paper made use of the FIR $Herschel$ images at 70, 160, 250, 350 and 500 $\mu$m from the $Herschel$ infrared Galactic plane survey (Hi-GAL). $Herschel$ is an ESA space observatory with science instruments provided by European-led Principal Investigator consortia and with important participation from NASA. 
The paper made use of the 1.1 mm dust continuum data from the Bolocam Galactic Plane Survey (BGPS). 
The paper made use of the 21 cm radio continuum data from the VLA Galactic Plane Survey (VGPS) and high resolution 20 cm radio continuum data from the Multi-Array Galactic Plane Imaging Survey (MAGPIS). 
The paper made use of the Galactic Ring Survey (GRS) $^{13}$CO data which conducted by the Boston University and the Five College Radio Astronomy Observatory (FCRAO). 
The paper also made use of the 6.7 GHz methanol maser data from the online tool $MaserDB$.

\section*{Data Availability}

The data supporting the plots within this paper are available on reasonable request to the corresponding author.
The UKIDSS data is available through a fully queryable user interface WSA (\url{http://wsa.roe.ac.uk/index.html}). The $Spitzer$ data and $Herschel$ data are available at IRSA (\url{https://irsa.ipac.caltech.edu/frontpage/}). The BGPS 1.1 mm data is available at CSO (\url{http://www.cso.caltech.edu/bolocam/}). The MAGPIS data can be obtained at \url{https://third.ucllnl.org/gps/}. The GRS $^{13}$CO data is available at \url{https://www.bu.edu/galacticring/new_data.html}. The 6.7 GHz methanol maser data can be retrieved from the online tool $MaserDB$ (\url{https://maserdb.net}).



\bibliographystyle{mnras}
\bibliography{N59bib} 

\bsp	
\label{lastpage}
\end{document}